# Tailored Thin Films: Modulating Soft Photonics with Dynamically Tunable Large Area Microstructures via Controlled Thermal Processing

*Srijeeta Biswas, Renu Raman Sahu, Omkar Deokinandan Nayak Shinkre, Shubham Meena, Ramnishanth, Mark Vailshery, Tapajyoti Das Gupta[*]*

Laboratory of Advanced Nanostructures for Photonics and Electronics, Department of Instrumentation and Applied physics, Indian Institute of Science, C.V. Raman Road, Bengaluru-560012, India

E-mail: tapajyoti@iisc.ac.in

Self-assembled nano and micro-structures, particularly those capable of responsive erasure and regeneration, have garnered significant interest for their applications in smart photonics and electronics. However, current techniques for modulating these architectures largely depend on network rearrangement, posing challenges for in situ regeneration. Furthermore, their common fabrication techniques are complex and uncontrolled with the structures formed not being amenable for large area applications, thus compromising their economic viability. Herein, we present a controlled thermal process strategy for fabricating large-area, dynamically tunable, 1D, 2D and 3D micro and nanostructures on a wide range of compatible materials including metals, semiconductors and polymers. By tuning the temperature changes in the system, thermal expansion coefficients of thin films and substrates, surface energy, Young's modulus and thickness of the thin films we achieve robust, uniform, periodic structures over extensive areas on soft and stretchable substrates. The process is further supported by a theoretical model that we developed and validated by experiments and simulations. To showcase the robustness of our approach, we present prototypes of dynamically tunable diffraction gratings, optical diffusers, large-area reflective displays, camouflage devices, out-coupling efficiency enhancers, wearable devices and mechanochromic sensors.

## 1. Introduction

Nano and micro-structures enhance light-matter interaction, thus enabling manipulation of electromagnetic radiation properties for tailored photonic device development[1–5]. These devices serve a wide array of purposes in healthcare[6–8], energy harvesting[9,10], smart displays[11–14], computation[15–18], communication[16], and wearable technology[19–22]. Thus far however, conventional fabrication methods rely on lithography and precise etching for each device, resulting in increased complexity, time consumption, and cost, hence hindering their scalability and cost effectiveness.

Non-lithographic techniques involving various kinds of exfoliation methods[23–26], microfluidics[13,27] and nanoimprint-lithography[19,28–30] involve multiple complex steps leading to fabrication challenges which again reduces the scalability and economic viability. Mechanical exfoliation requires precise peeling of uniform thin films without causing any damage to the micro-structures, microfluidics hinges upon incorporation of microfluidic channels that are difficult to design and fabricate over large areas, nanoimprint-lithography is often plagued with defects arising from resist material inconsistencies and challenges with overlay alignment. Photonic devices which are attainable through diverse self-ordered surface morphologies such as spikes, wrinkles, and folds on substrates[31,32] could potentially solve these challenges. Wrinkled surfaces having well-ordered periodicities can act as meta-surfaces for modulating light matter interactions. Various techniques, like photoinduced methods involving LASER[33–35] and UV[10,36–38] treatments, chemical swelling[38–41], mechanical pre-straining[33,42,43], and thermal annealing[35,44,45], have been utilized to create these wrinkles. Methods like mechanical straining, thermal annealing, and UVO/plasma treatments cover extensive areas, however, they lead to multiple cracks that disrupt the continuity of the ordered wrinkles and hence are not suitable for development of large area photonic devices[46–51]. Alternate techniques, such as Laser treatments and chemical swelling exhibiting fewer defects are limited to smaller area and tend to produce disorganized patterns[34,41,52–54]. Thus far, however, much of the existing work focuses on irregular and disorganized structures or on methods either limited by their area or plagued by multiple defects such as cracks.

Herein, we develop a simple yet scalable thermal processing technique for fabricating large area, dynamically tunable, customized 1D, 2D and 3D wrinkles. By leveraging on the control of the thermal expansion coefficient mismatch of the film and the substrate, thickness, surface energy of the thin film and the change in temperature of the system we show a regulated control over the structure, orientation, and periodicity of the wrinkles. Further, to understand the process we introduce a universal mathematical model to verify the experimental results and to calibrate the process to achieve wrinkles of desired periodicity and orientation.

In order to demonstrate the effectiveness and versatility of our process, we showcase prototypes of various photonic devices such as tunable temperature-controlled diffraction gratings, optical diffusers, body-motion sensors, dynamically tunable camouflage devices and flexible reflective displays, that can find applications in healthcare and smart sensors. The performance of the devices is consistently repeatable over 100s of cycles of thermal and mechanical stimuli, thus confirming the high fidelity, reliability and robustness of the samples prepared through our processing technique. The samples have also shown promising results for increased outcoupling efficiencies for encapsulated light sources and thus can find potential applications in energy harvesting devices like solar cells.

## 2. Fabrication of ordered wrinkled structures

Our fabrication process, as illustrated in Figure 1a and elaborated in the Methods section, Supplementary Information section 1 and Supplementary Figure 1, involves the use of bilayer system consisting of thin films deposited on polydimethylsiloxane (PDMS) substrates with varying degrees of softness (Supplementary Figure 2d). PDMSx is prepared by combining x grams of base (Sylgard 184, Dow Corning) with 1 gram of curing agent. The ratio of curing agent determines the substrate's softness (Supplementary Figure 2d) and its thermal expansion coefficient (Supplementary Information section 1), a higher proportion results in lower softness and lower thermal expansion coefficient. Thin films of chalcogenide (for example $As_2Se_3$), metals (like Ag or Al), or polymer (like PMMA) are then deposited via simple physical deposition techniques (thermal evaporation, sputtering and by spin coating) onto the cured PDMS substrates, followed by thermal processing methods involving simple heating/joule heating and subsequent cooling to create ordered structures.

Figure 1a presents the schematic and SEM image of the sample after thin film deposition and prior to thermal processing, along with the schematic and SEM image following thermal processing, during which the film self-assembles into periodic wrinkled structures. This self-assembly due to thin film surface instabilities occurs for different film thicknesses as shown in Figure 1b, thus making it apparent that the film thickness plays a pivotal role in determining the periodicity or the wavelength of the structures.

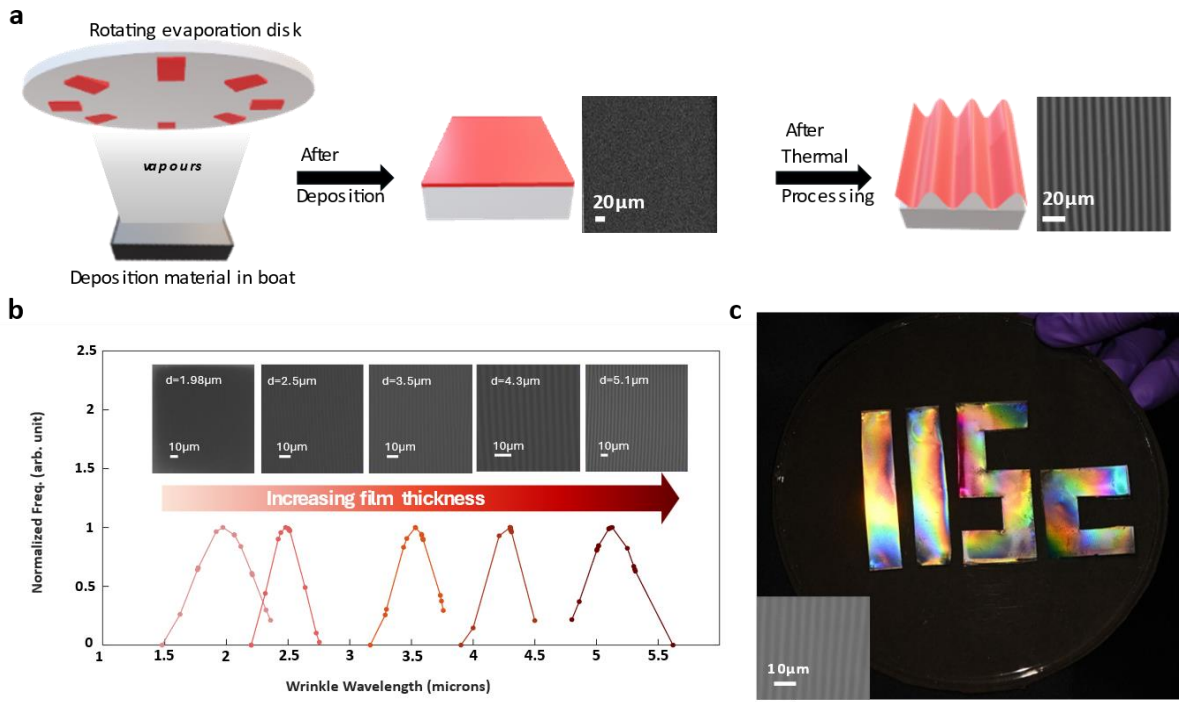

*Figure 1*. Fabrication and top view morphology of wrinkled samples. a| Schematic and corresponding SEM images of thermal evaporation of $As_2Se_3$ onto PDMS substrate and subsequent thermal processing to obtain the desired wrinkled structures. b| Wrinkle wavelength size distribution obtained with thermal processing of $As_2Se_3$ deposited on PDMS 5 substrate, showing the tunability of wrinkle wavelength. Inset: Corresponding SEM images for different film thicknesses, d is the periodicity of the wrinkles. The samples were heated to 80°C before cooling. c| Optical photograph showing a large area (8cm×6cm) IISc logo-shaped wrinkled surface acting as a flexible reflective display. Inset: SEM image of the corresponding wrinkled structure. Material: Ag on PDMS 5.

Our approach is simple and scalable, as evident from Figure 1c wherein we fabricate a large area(8cm×6cm) IISc logo consisting of periodic grating structure acting as a reflective display consisting of Ag thin film deposited on PDMS 5 substrate. The inset of Fig. 1c shows an SEM view of the patterned area with a nearly perfect array of micro-plasmonic lines.

## 3. Attainable morphologies and choice of materials

Through our single-step thermal processing approach, we've achieved a diverse array of morphologies by adjusting various parameters like the thermal expansion coefficients of the film and substrate, film thickness, and system temperature. The AFM images of the various morphologies are shown in Figures 2a-e, a: nano cones, b: 1D sinusoidal wrinkles, c: 2D zigzag wrinkle, d: isotropic labyrinth wrinkles superimposed over 1D sinusoidal wrinkles, e: checkerboard wrinkles. Our approach accommodates various materials, such as chalcogenides like $As_2Se_3$(Figure 2f), metals like aluminum (Figure 2g) and silver (Figure 2h), and polymers like PMMA (Figure 2i), thereby broadening the scope of applications across multiple areas from all dielectric, through low index polymers as well as plasmonic structures. Figure 2j shows SEM image of large area ($1200\ \mu m \times 1200\ \mu m$) ordered 1D sinusoidal wrinkles.

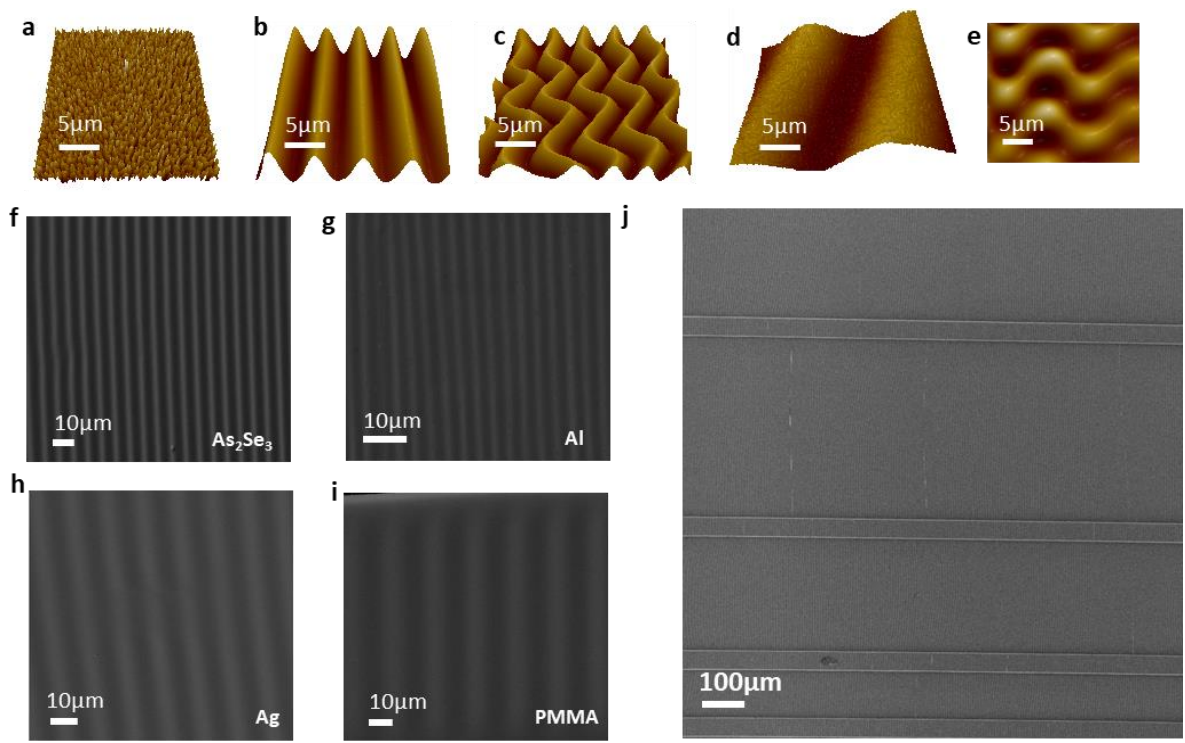

*Figure 2.* Attainable morphologies and compatibility with different materials. a-e| AFM images of the various morphologies attained. Material: $As_2Se_3$ on PDMS substrate. a| nano cones, b| 1D sinusoidal wrinkles, c| 2D zigzag wrinkles, d| isotropic labyrinth wrinkles on 1D sinusoidal wrinkles, e| Checkerboard wrinkles. f-i| SEM images of wrinkles achieved on various materials through our process. Substrate: PDMS. f| $As_2Se_3$, g| Aluminum, h| Silver, i| PMMA. j| SEM image of large area ordered 1D sinusoidal wrinkle. Area: ~1200µm×1200µm. Material: $As_2Se_3$ on PDMS substrate

### 3.1. Role of substrate properties: thermal expansion coefficient and refractive index

The formation of wrinkles necessitates substrates with an ideal thermal expansion coefficient. Rigid substrates such as glass and silicon, which possess low coefficients of thermal expansion, experience minimal expansion and contraction during heating and cooling, respectively. Consequently, they lack the capacity to induce the necessary level of contraction in thin films crucial for wrinkle formation (Supplementary Figure 2a, b). Conversely, substrates such as PDMS 15 and above, which have extremely high coefficients of thermal expansion (Supplementary Information section 1.2), expand significantly, resulting in the cracking and subsequent dewetting of thin films deposited atop them (Supplementary Figure 2). PDMS 3, 5, 10 have been found to possess the optimum thermal expansion coefficient that is required for wrinkle formation. Additionally, the substrates must have a minimum thickness such that they can act as a semi-infinite system. Below this thickness they will undergo global buckling upon thermal treatment, thereby hampering the formation of uniform wrinkles. In addition, PDMS's transparency and non-absorption within the visible and the NIR (Near Infrared) spectra render it highly promising for photonic applications in these regions.

## 3.2. Role of thin film properties: thermal expansion coefficient, film thickness, surface energy

In addition to the optimal thermal expansion coefficient of the substrate, the formation of wrinkles also hinges on an ideal thermal expansion coefficient of the deposited thin film. If this coefficient is very high or similar to that of the substrate, the contraction force experienced during cooling is not adequate for wrinkle formation (Supplementary Figure 2a, b). Conversely, if the coefficient is excessively low, then the film will experience very high tensile stress during heating which will lead to the formation of cracks, ultimately favoring dewetting over wrinkle formation in achieving equilibrium. Within the optimal range of difference in thermal expansion coefficients of the substrate and the film, a greater difference between them results in the formation of wrinkles with larger wavelengths. This relationship is elucidated in the modeling section in Supplementary Information section 2.3.2 and illustrated in Supplementary Figure 9. For a given substrate and thin film material, wrinkle formation requires a minimum film thickness, contingent upon the temperature change within the system and surface energy of the film (Supplementary information section 2.3.3). The lower the surface energy, the lower the required minimum film thickness (illustrated in Supplementary Figure 11). Higher film thickness results in a greater stress energy in the film during cooling. When combining lower surface energy with higher film thickness, the resulting wrinkles have longer wavelengths. These phenomena are elaborated upon in the modeling section.

## 3.3. Role of temperature

The formation of large area, uniform, ordered wrinkles requires a uniform force acting on the film which can be achieved within an optimal temperature range. The lack of wrinkle formation at temperatures much below this range is attributed to the insufficient generation of compressive stress on the film (Supplementary information section 2.6.1 and Supplementary Figure 14). Conversely, when the system is heated to temperatures exceeding this range by a large extent, disorderedness starts in the thin films (Supplementary information section 2.6.1 and Supplementary Figure 15). At temperatures just below this optimal range, wrinkles with non-uniform amplitudes are formed due to the non-uniformity of the force acting on the film. It is highest near the center of the film and gradually diminishes towards both sides. Similarly at temperatures just above the optimal range, the system experiences global buckling and wrinkle bifurcation which again leads to non-uniformity in the wrinkles. In our case, we have experimentally determined the optimal range to be between 80°C to 120°C for $As_2Se_3$ thin films deposited on PDMS 5 and 10 substrates.

## 4. Wrinkle formation on PDMS substrates

Due to the thermal expansion coefficient mismatch between the film and the substrate (as included in the Supplementary Information section 1.2), when the system is heated, the substrate expands more than the film. Consequently, the film experiences a tensile stress. During subsequent cooling, the substrate contracts more than the film. Again, in order to maintain compatibility, the film experiences compressive stress while the substrate experiences tensile stress. To counteract these unbalanced stresses and moments, wrinkles are formed in the film (Supplementary Figures 3 and 4). Thus, there exists a critical wrinkle wavelength that minimizes the total strain energy in the system which is dependent on the material properties of both the film, substrate and the temperature change within the system. The stress acting on the film is then defined as

$$\sigma = Y_f(\alpha_s - \alpha_f)(T_f - T_i) \qquad (1)$$

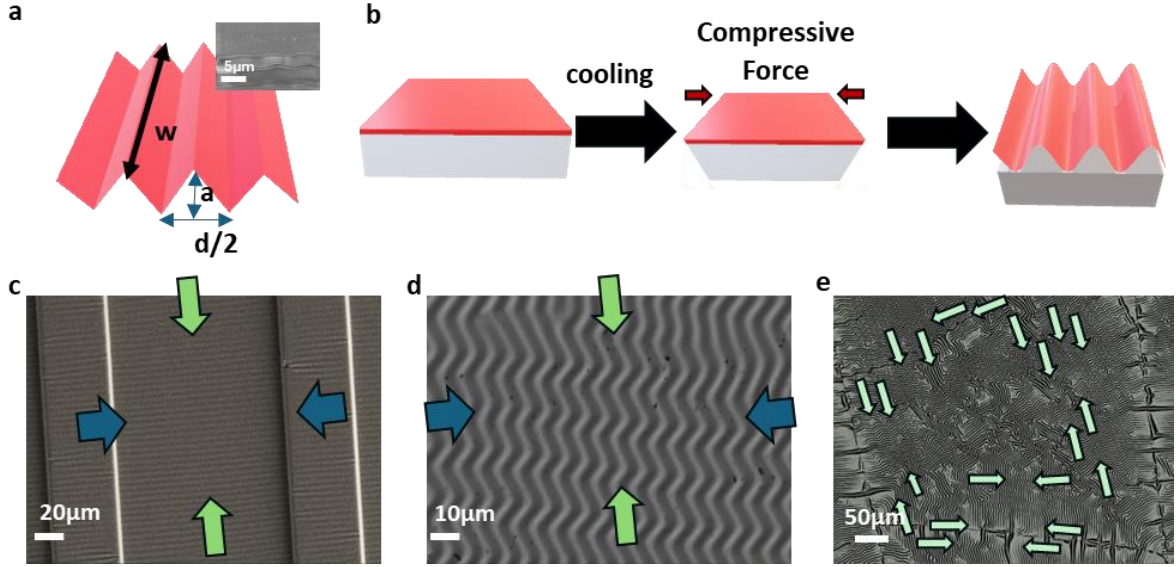

***Figure 3.*** *Explanation of wrinkle formation mechanism. a| Schematic of 1D sinusoidal wrinkle (assumed to be triangular for modelling purpose). w is the width, a is the amplitude and d is the periodicity i.e. the wavelength of the wrinkle. b| Schematic showing the cooling process leading to the formation of wrinkles. c-e| SEM images showing various wrinkle patterns. c| 1D sinusoidal wrinkles, d| 2D zigzag wrinkles, e| isotropic labyrinth wrinkles. Blue arrow: direction of lateral compressive stress on film. Green arrow: direction of longitudinal compressive stress on film. Light-green arrow: direction of isotropic compressive stress on film.*

where $Y_f$ is the Young's modulus of the film, $\alpha_s$ and $\alpha_f$ are the thermal expansion coefficient of the substrate and the film respectively. $T_f$ and $T_i$ are the final and initial temperature of the film during heating or cooling. When $T_f$ is greater than $T_i$ i.e. the system is being heated, the stress acting on the film is tensile, whereas when $T_f$ is less than $T_i$ i.e. the system is being cooled, the stress acting on the film is compressive. Here we have considered the Young's modulus of the film and not its plane strain modulus because the wrinkle formation in the film is solely due to the compressive stress experienced by it due to the mismatch in the thermal expansion coefficient and not due to Poisson effect. (Illustrated in Supplementary video 1).

## 4.1. Modeling: Wrinkle formation through controlled thermal processing

To understand the mechanisms behind wrinkle formation, we developed a mathematical model that reproduces the following trends that are observed experimentally:

a. Wrinkle wavelength increases with increase in film thickness
b. Higher substrate softness (i.e. higher thermal expansion coefficient) leads to the formation of wrinkles with higher wavelength.
c. Thin films with higher surface energy form wrinkles with lower wavelengths.
d. Higher temperature during heating leads to an increase in the wavelength of the wrinkles.
e. Below a critical length along the direction of compressive force acting on the film, wrinkles are not formed.

The periodicity of the wrinkles is given as[32,55,56],

$$d = 2\pi h \left(\frac{E_f}{3E_s}\right)^{1/3}$$

Here $E_f$ and $E_s$ represent the plane strain modulus of the film and the substrate, while h denotes the thickness of the film deposited on the substrate, and d signifies the periodicity of the wrinkles. In our thermal processing method, the determining factors are the thermal expansion coefficients of both the film and the substrate, the thickness and surface energy of the thin film, and the temperature variation within the system. Figure 3a depicts the schematic of the wrinkled structure and cross-sectional FIB image of the bilayer system. For modeling purposes, we assume the sinusoidal wrinkles to be triangular in nature. When the bilayer system consisting of a rigid uniform thin film deposited over a flexible substrate undergoes heating above the room temperature followed by cooling back to the room temperature, the substrate contracts more than the film thereby exerting a compressive stress over the latter. Wrinkle formation takes place to counteract this compressive stress on the film. Figure 3b shows the schematic illustration of the wrinkle formation process (Wrinkle formation mechanism is explained in detail in Supplementary Information section 2.2). At the verge of wrinkle formation, the rate at which the surface energy increases (due to the formation of new surfaces i.e. the two inner slopes of each sinusoidal wrinkle) is delicately balanced by the elastic strain energy i.e. the compressive strain energy. This equilibrium yields our wavelength equation, through the equalization of these energies

The equalization of the energies leads to the relation

$$d = \frac{Y_f \, [(\alpha_s - \alpha_f)\Delta T]^2 hL}{Cn\gamma_f} \qquad (3)$$

Here $d$ is the periodicity(wavelength) of the wrinkles, $\alpha_s$ and $\alpha_f$ are the thermal expansion coefficients of the substrate and the film respectively, $\Delta T$ is the change in temperature of the system during the cooling process, $C$ is a proportionality constant, greater than 1(details in Supplementary Information section 2.2). $Y_f$, $\gamma_f$, $h$ are the Young's modulus, surface energy and thickness of the deposited film respectively. $L$ is the expanded length of the film after heating, along which the compressive force starts acting, and $n$ is the number of wrinkles formed within the length $L$. From this relation we get the final Wavelength equation which is defined as

$$d = C_w \frac{Y_f \, [(\alpha_s - \alpha_f)\Delta T]^2 h}{\gamma_f} \qquad (4)$$

Where $C_w$ is a proportionality constant depending on certain parameters and having the units of length (detailed in Supplementary Information section 2.2).

This equation enables the prediction of the wavelength of wrinkles formed in a thin film, based on its known surface energy, Young's modulus, thickness, and thermal expansion coefficient mismatch, for variations in temperature. Figure 4a shows that as ΔT(illustrated in Supplementary video 1), film thickness(h) and thermal expansion coefficient of substrate($\alpha_s$) increases, wavelength of wrinkles also increases in accordance with our mathematical model (values of $C_w$ considered for matching our Wavelength Equation plots with the corresponding experimental data have been detailed in Supplementary information section 2.3.2 and Supplementary Figure 9).

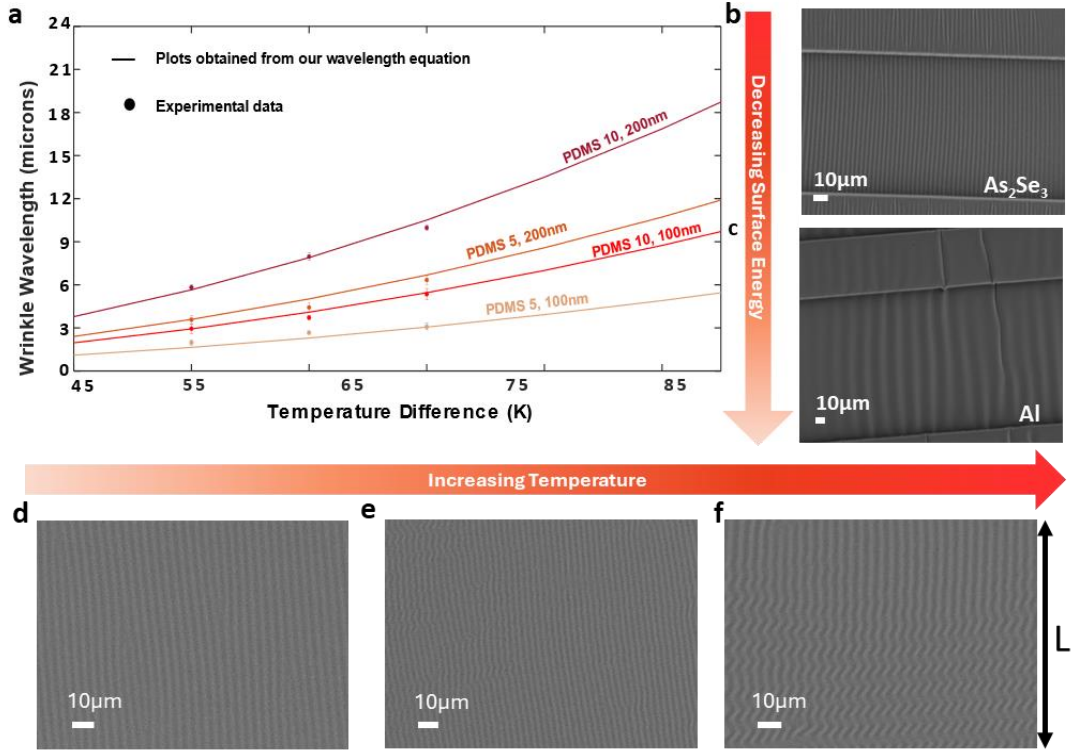

*Figure 4.* Experimental verifications of our mathematical model. a| Wavelength of wrinkles vs. $\Delta T$ plot. The solid circles depict the experimental data, and the solid lines are their corresponding plots derived from our Wavelength Equation. Material: $As_2Se_3$ on PDMS substrate. It shows that as $\Delta T$, film thickness($h$) and thermal expansion coefficient of substrate($\alpha_s$) increases, wavelength of wrinkles also increases, which verifies our mathematical model. b-c| Under exactly similar conditions, 100nm thin film deposited on PDMS 5 substrate and heated to 90°C, Al film has much larger wavelength of wrinkles (~20 microns) as compared to that of $As_2Se_3$ (~3 microns) since the surface energy Al is much less than that of $As_2Se_3$. d-f| Illustrates the dependence of critical length $L$ on $\Delta T$. It shows that as $\Delta T$ increases the critical length needed for wrinkle formation along that direction decreases. d| At $\Delta T=75°C$, there's no wrinkle formation along the direction of L and we have 1D sinusoidal wrinkles. e| At $\Delta T=105°C$, L is almost equal to the critical length needed for wrinkle formation along that direction. f| At $\Delta T=115°C$, wrinkles have also formed along the direction of L, and we have 2D sinusoidal wrinkles.

Figures 4 b-c shows the dependence of the wavelength of the wrinkles on the surface energy of the thin film material. $As_2Se_3$ has a surface energy of $0.2J/m^2$ and that of Al is $0.0396J/m^2$. Thus, for 100nm thin films deposited on PDMS 5 and heated to 90°C, the wavelength of the wrinkles formed is inversely proportional to the surface energy of the thin film. In Figure 4b the wavelength of the wrinkles is 2.9μm and in Figure 4c it is 20.3μm. Figures 4d-f Illustrates the dependence of critical length L on $\Delta T$. It shows that as $\Delta T$ increases the critical length needed for wrinkle formation along that direction decreases. At $\Delta T=75°C$, there's no wrinkle formation along the direction of L and we have 1D sinusoidal wrinkles. At $\Delta T=105°C$, L is almost equal to the critical length needed for wrinkle formation along that direction. At $\Delta T=115°C$, wrinkles have also formed along the direction of L, and we have 2D sinusoidal wrinkles. Note that, to witness wrinkle formation, the temperature to which the system is heated should be greater than the critical temperature, as discussed in the previous section. Our mathematical model can also be effectively used to find the Thermal expansion coefficients of thin films. (Illustrated in Supplementary Information section 2.4).

## 4.2. Role of ridges on the substrate

Ridges are made on the PDMS substrates (details in methods) to control the orientation of the wrinkles. The distance between them varies from 150 microns to 950 microns in steps of 200 microns. In order to form wrinkles in a particular direction, the compressive stress along that direction should exceed the critical stress needed for wrinkle formation. The compressive stress acting on the film is defined as (detail derivation in Supplementary Information section 2.2)

$$\sigma = E_f[(\alpha_s - \alpha_f)\Delta T]^2 whL \qquad (5)$$

Therefore, it becomes apparent that the strain energy in a particular direction is directly correlated with the length it acts upon. This necessitates the requirement for a critical length for wrinkle formation along a given direction. We derive our Critical Length Equation from Equation 3 by analyzing the equilibrium condition and is defined as (detail derivation and explanation in Supplementary Information section 2.5)

$$L_c = C_l \frac{\gamma_f d}{Y_f h[(\alpha_s - \alpha_f)\Delta T]^2} \qquad (6)$$

Where $L_c$ is the critical length. $C_l$ is a unitless proportionality constant and is $= Cn$. (detailed in Supplementary information section 2.5). The SEM images presented in Figure 3c-e depict various orientations of wrinkles. In Figure 3c, the wrinkles exhibit a 1D sinusoidal order. In this scenario, the longitudinal compressive stress surpasses the critical stress, while the lateral stress falls below it. Consequently, wrinkles develop solely along one direction. In Figure 3d, the wrinkles adopt a 2D sinusoidal pattern. Here, both longitudinal and lateral compressive stresses exceed the critical stress; however, the longitudinal stress prevails, resulting in longer wavelengths along this direction compared to the lateral one. Finally, in Figure 3e, the wrinkles form a labyrinthine isotropic pattern. At that instance, both longitudinal and lateral compressive stresses are comparable and surpass the critical stress. Adjusting the distance between the ridges allows for modulation of the lateral compressive stress and, consequently, the orientation of the wrinkles.

## 4.3. Effect of heating temperature and PDMS curing time on wrinkle formation and critical length

As the curing time of the PDMS substrate increases, the crosslinking between the PDMS base and the curing agent also increases, thereby reducing the softness of the substrates. With decrease in substrate softness, its thermal expansion coefficient also decreases leading to a decrease in the thermal coefficient mismatch between the film and the substrate. This reduces the wrinkle wavelength (Supplementary Figure 10). Again, as the substrate curing time increases, the wrinkles become more ordered with respect to increase in heating temperature as illustrated in Supplementary Information section 2.6.1 and SI Figure 15).

## 5. Optical properties and their dynamic tuning

The wavelength of the wrinkles impacts the optical properties. The wrinkles act as sinusoidal diffraction gratings[42,43,57,58] and their spacing (wavelength) and orientation determines the diffraction pattern formed by them. For 1D sinusoidal wrinkles we get 1D diffraction spots, the diffraction order depending on the periodicity of the wrinkles.

For an incident light of wavelength λ, incident at an angle $\theta_i$ , on a sinusoidal grating of periodicity d (wavelength of the wrinkles), the path difference between the incident light on the adjacent wrinkle crests is $d \sin \theta_i$. Similarly, the path difference between the reflected light from the adjacent wrinkle crests is $-d \sin \theta_d$, where $\theta_d$ is the angle at which the rays are detected. Thus, we come to the diffraction equation

$$m\lambda = d(\sin \theta_i + \sin \theta_d) \qquad (7)$$

Where m is the diffraction order of the light of wavelength $\lambda$ . Henceforth, it becomes evident that altering the spacing of wrinkles while maintaining all other parameters constant results in a variation in the diffraction order. By varying the temperature of the system, we can dynamically tune the periodicity of the wrinkles, thereby the diffraction order (Figure 6a, Supplementary Figure 19). When the incident beam aligns parallel to the direction of the wrinkles, vertical diffraction spots appear, while aligning it perpendicular to the wrinkles produces horizontal diffraction spots. In the presence of isotropic labyrinth wrinkles, circular diffraction patterns are observed. (Supplementary Information section 3.2). Figures 5a, 5b, and 5c depict schematics illustrating the orientation of the incident beam plane relative to the direction of the wrinkles. In Figure 5a, the incident beam plane intersects parallel to the wrinkle direction, resulting in a vertical array of diffraction spots, as depicted in Figures 5d(top) and 5e(top). Conversely, in Figure 5b, the incident beam plane aligns perpendicular to the wrinkle direction, generating a horizontal array of diffraction spots, as illustrated in Figures 5d(middle) and 5e(middle). Figures 5d and 5e represent diffraction spots for the same sample, where a transition from green to red incident beam reduces the number of diffraction spots (diffraction order). In Figures 5d (top, middle) and Fig.5e (top, middle) the experimental images are on the left and the corresponding simulation images are on the right. Since the structures formed have a one-dimensional sinusoidally varying amplitude, to get the simulation results we first calculate the phase in a plane near the structure immediately after scattering. From this phase, we derive the field and propagate it using scalar diffraction theory to produce the output. (details in Supplementary Information section 3.6). In Figure 5c, the beam interacts with isotropic labyrinth wrinkles, yielding diffraction rings, as shown in Figures 5d(bottom) and 5e(bottom). The wrinkled structures can also serve as reflection-based optical diffusers. When the samples are heated above 140°C and then allowed to cool, the number of diffraction rings increases continuously until the sample reaches room temperature (Supplementary Information section 3.3).

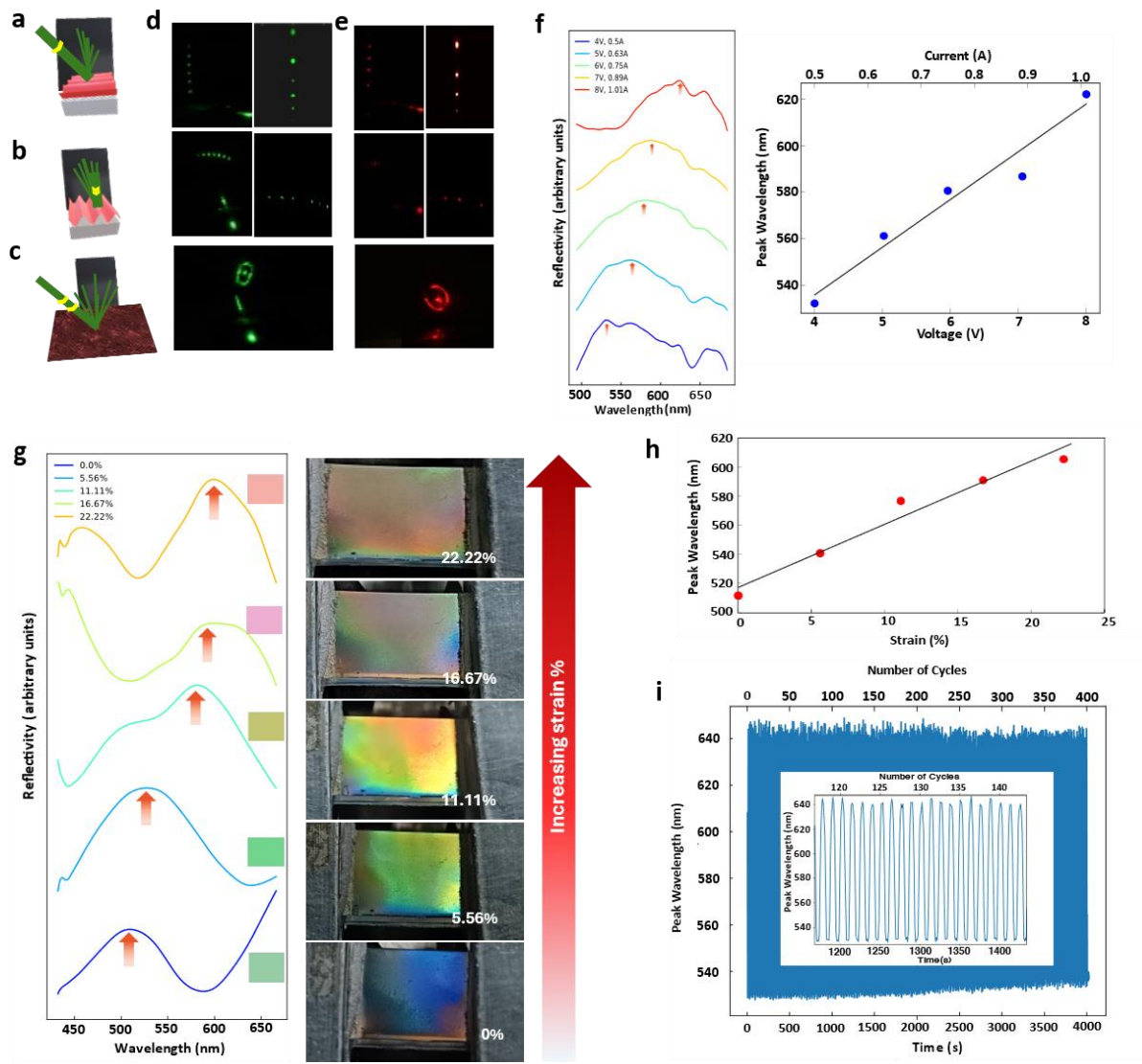

***Figure 5.*** *Optical properties of wrinkles. a| Schematic showing the incident beam plane intersecting parallel to the wrinkle direction, resulting in a vertical array of diffraction spots. b| Schematic showing the incident beam plane intersecting perpendicular to the wrinkle direction, resulting in a horizontal array of diffraction spots. c| Schematic showing the incident beam interacting with isotropic labyrinth wrinkles, resulting in diffraction rings. (d) and (e) depict the experimental(left) and corresponding simulation results(right) for incident red and green lights for the adjacent orientations illustrated in (a), (b) and (c). f|(Left) Reflectivity spectrum obtained experimentally as the sample was heated through joule heating and subsequently cooled. (Right) Peak wavelength of reflectivity vs. Current-Voltage plot showing a sensitivity of 20nm/V. With increase in current passing through the wires, a red shift is observed in the peak wavelength during the cooling process. The spectrometer detector was placed at an angle of 35° with respect to the sample and a broad white light source was placed above the sample. g| Left: Reflectivity spectrum obtained experimentally as the sample was stretched uniaxially from 0% to 22% along the direction of wrinkles. The spectrometer detector was placed at an angle of 35° with respect to the sample and a broad white light source was placed above the sample. The CIE colours represented by each plot are shown in insets. Right: Images of corresponding samples captured with a camera placed at an angle of 35° and at a radius of 10cm with the sample as the center. h| Peak wavelength of reflectivity vs. Strain % plot. i| Peak reflectivity vs. time & Number of cycles for a periodic uniaxial linear strain for 400 cycles.*

Figure 5f(left) depicts the reflectivity spectrum obtained experimentally as the sample is heated through joule heating and subsequently cooled. With the increase in current passing through the heating wires, a red shift is observed during the subsequent cooling process due to an increase in the periodicity of wrinkles (experimental details in Methods section). Figure 5f(right) illustrates the Peak wavelength of reflectivity vs. Current-Voltage plot having a sensitivity of 20nm/V. Figure 5g and Supplementary video 2 show a red shift in the color of the sample with increase in strain. By applying a uniaxial mechanical strain along the direction of wrinkles, the periodicity of the wrinkles increases. For a constant $\theta_i$ and $\theta_d$ as the periodicity of the wrinkles increases, the wavelength that is getting reflected in a particular direction also increases, thereby causing the red shift. Each successive reflection spectra and the adjoining sample image corresponds to a 1mm increase in the length of the sample. The samples (Figure 5g) consist of wrinkled Ag thin films deposited over PDMS 5. On decreasing the strain, one observes a blue shift in the reflection spectrum. The peak wavelength of reflection vs. strain% plot shown in Figure 5h depicts a shift in spectral features by about 30nm per millimeter of sample length change. The color saturates after a strain of around 25%. To show the repeatability and robustness of the spectral characteristics, the samples were cycled for a minimum of 400 cycles as shown in Figure 5i, thus establishing it as a reliable, long-lasting mechanochromic sensor.

## 6. Applications

The wrinkles can be tuned dynamically through temperature changes (Supplementary video 1). Thus, by changing the heating temperature of the sample and subsequently cooling it down to room temperature we can get different diffraction orders upon shining monochromatic light over it. Figure 6a shows the different diffraction orders as the system is heated to different temperatures and cooled down subsequently. The sample is made of 150nm $As_2Se_3$ thin film deposited atop PDMS 5 substrate. As the temperature to which the sample is heated increases, the wavelength of the wrinkles formed on subsequent cooling also increases, thereby increasing the number of diffraction orders. Simulation results for the respective experiments are shown in insets. This process of heating to higher temperatures and subsequent cooling is repeatable upto more than 50 times (Supplementary Information section 3.5, Supplementary Figure 19).

The presence of wrinkles increases the angle of the total internal reflection cone, thereby increasing the outcoupling efficiency[59] (detailed explanation in Supplementary Information section 4 and Supplementary Figure 21). When an LED is encapsulated by a wrinkled surface consisting of transparent PMMA wrinkles on PDMS substrate, the detector registers a power of 3.526mW, whereas when it is encapsulated by plain PDMS, the detector detects a power of 2.779mW (Figure 6b). Figure 6c shows the simulation result for the above experiment. Wrinkled surface shows a transmittance of 98% as compared to a transmittance of around 87% when encapsulated by an unpatterned PDMS substrate.

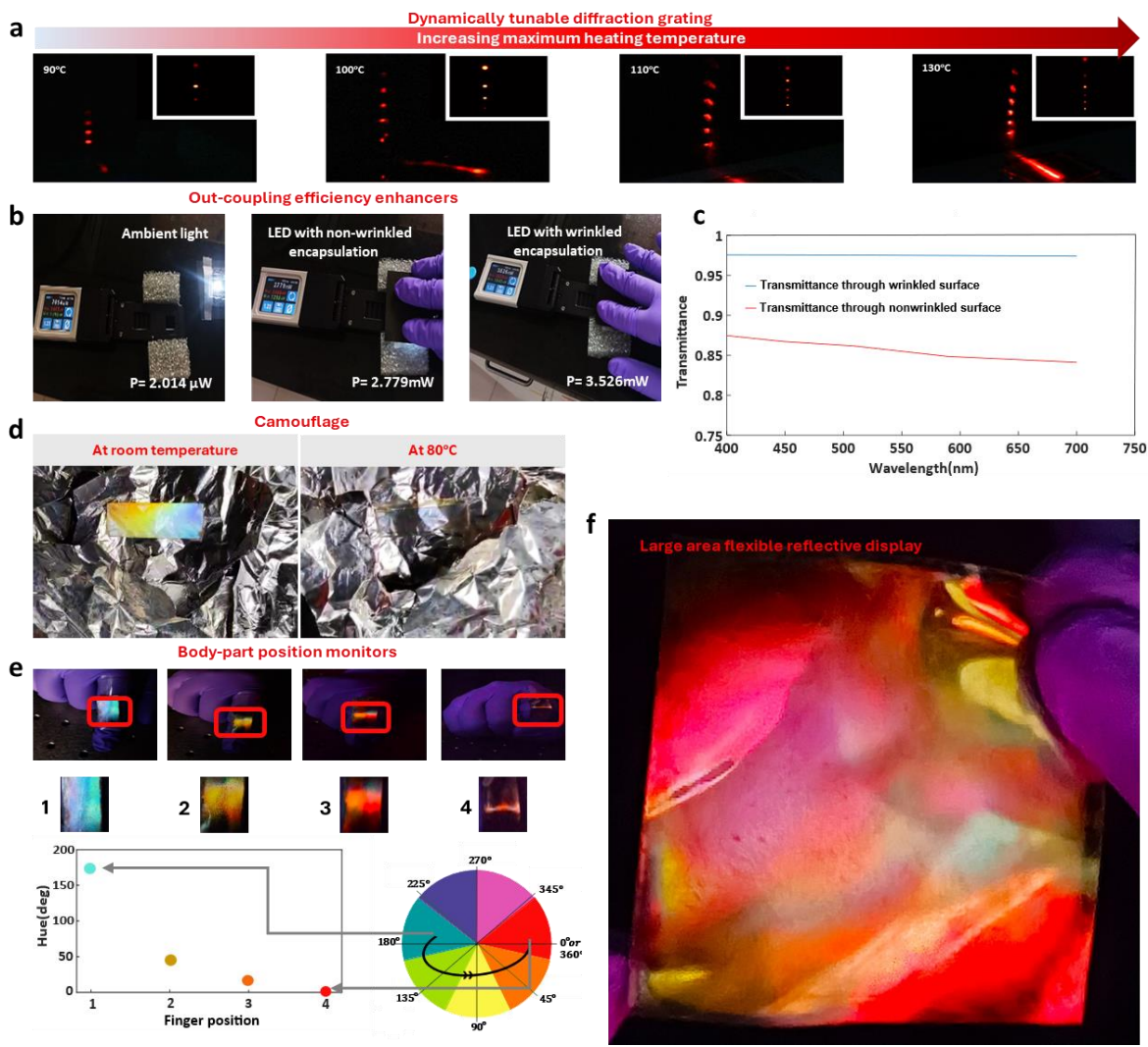

***Figure 6.*** *Applications of wrinkles a| Dynamically tunable temperature-controlled diffraction grating. Experimental results of diffraction patterns as the system is heated to different temperatures and cooled down subsequently. Material: 150nm $As_2Se_3$ on PDMS5. Insets show the simulation results for the respective experiments. b| Increase in out-coupling efficiency of encapsulated light sources. LED lights encapsulated by transparent wrinkles (PMMA wrinkles on PDMS substrate) register higher power as compared to those encapsulated by flat unpatterned PDMS. When the LED is encapsulated by a wrinkled surface, the detector registers a power of 3.526mW, whereas when it is encapsulated by plain PDMS, the detector detects a power of 2.779mW. c| shows the simulation result for the above experiment. Wrinkled surface shows a transmittance of 98% as compared to a transmittance of around 87% when encapsulated by an unpatterned substrate. d| Adaptive thermoresponsive visible camouflage. Colors disappear on being heated and reappear on being cooled. e| Mechanochromic body-part position monitor. f| Large area angle-dependent flexible reflective display.*

The isotropic labyrinthine wrinkles give rise to circular diffraction rings, thus acting as temperature controlled tunable optical diffusers. The details have been explained in Supplementary Information section 3.2, 3.3 and visually represented in Supplementary Figure 18.

Figure 6d demonstrates that the samples can act as thermally responsive camouflage (Supplementary video 3). At room temperature, the presence of the wrinkles leads to the diffraction of any light falling on it thereby making them visible. When heated to higher temperatures such that the wrinkles vanish, the samples become invisible in a grey background (in case of Ag/Al coated samples).

Figure 6e demonstrates the prototype of a body motion sensor. On increasing the curvature of the finger, the stress acting on the bilayer system increases which in turn increases the periodicity of the wrinkles and a red shift is observed in the color of the sample on shining white light over it. The adjacent Hue vs. finger position plot shows the red shift in sample color for increasing curvature

These samples can also be utilized to create large area, angle-dependent reflective displays. As illustrated in Figure 6f, an optical photograph demonstrates a flexible reflective display featuring a large area (6cm × 6cm) wrinkled surface. After deposition and before thermal processing the samples appear greyish (in case of Ag/Al coated samples). Upon thermal processing the samples give colors due to diffraction thereby confirming the role of wrinkles in the reflective displays as illustrated in Supplementary video 4.

## 7. Conclusion

By harnessing the disparity in thermal expansion coefficients between rigid thin films and flexible substrates, adjusting the spacing between ridges, and controlling the thermal processing approach, we have showcased a single-step and scalable fabrication method to generate ordered sinusoidal wrinkles. This method enables the production of devices for adaptive visible camouflage, dynamically tunable diffraction gratings of high repeatability, ring-like diffraction patterns, large area flexible reflective displays, wearable devices like body-motion sensor, substrates capable of enhancing the out-coupling efficiency of encapsulated light sources and energy harvesting devices such as solar cells.

The uniqueness of this method is that it's controlled and calibrated and can produce large area, defect-free, ordered wrinkles in a single step. It significantly decreases both the complexities involved in fabrication and the production costs associated with creating well-ordered micro-structures. This approach can also effectively ascertain the Young's modulus of thin films. Thanks to the spatial manipulation and control of wrinkles through stimuli like temperature, these smart surfaces could be utilized in applications such as smart displays, data storage, and anti-counterfeiting measures. Consequently, this bilayer design offers a promising approach for developing high-performance anti-counterfeiting materials that feature enhanced security, dynamic properties, and easy method of preparation.

## 8. References

1.  De Tommasi, E. *et al.* Frontiers of light manipulation in natural, metallic, and dielectric nanostructures. *Rivista del Nuovo Cimento* vol. 44 Preprint at https://doi.org/10.1007/s40766-021-00015-w (2021).

2.  Dupré, M., Hsu, L. & Kanté, B. On the design of random metasurface based devices. *Sci Rep* **8**, (2018).

3. Wang, W. & Qi, L. Light Management with Patterned Micro- and Nanostructure Arrays for Photocatalysis, Photovoltaics, and Optoelectronic and Optical Devices. *Advanced Functional Materials* vol. 29 Preprint at https://doi.org/10.1002/adfm.201807275 (2019).

4. Gupta, S. Single-order transmission diffraction gratings based on dispersion engineered all-dielectric metasurfaces. *Journal of the Optical Society of America A* **33**, 1641 (2016).

5. Wang, Q. *et al.* Optically reconfigurable metasurfaces and photonic devices based on phase change materials. *Nat Photonics* **10**, 60–65 (2016).

6. Zhang, S. *et al.* Metasurfaces for biomedical applications: Imaging and sensing from a nanophotonics perspective. *Nanophotonics* **10**, 259–293 (2021).

7. Yavas, O., Svedendahl, M., Dobosz, P., Sanz, V. & Quidant, R. On-a-chip Biosensing Based on All-Dielectric Nanoresonators. *Nano Lett* **17**, 4421–4426 (2017).

8. Xiong, R. *et al.* Biopolymeric photonic structures: Design, fabrication, and emerging applications. *Chemical Society Reviews* vol. 49 983–1031 Preprint at https://doi.org/10.1039/c8cs01007b (2020).

9. Sun, S. *et al.* Gradient-index meta-surfaces as a bridge linking propagating waves and surface waves. *Nat Mater* **11**, 426–431 (2012).

10. Kim, J. B. *et al.* Wrinkles and deep folds as photonic structures in photovoltaics. *Nat Photonics* **6**, 327–332 (2012).

11. Ma, Q. *et al.* Smart metasurface with self-adaptively reprogrammable functions. *Light Sci Appl* **8**, (2019).

12. Naveed, M. A. *et al.* Novel Spin-Decoupling Strategy in Liquid Crystal-Integrated Metasurfaces for Interactive Metadisplays. *Adv Opt Mater* **10**, (2022).

13. Zhu, Z. *et al.* Microfluidics-Assisted Assembly of Injectable Photonic Hydrogels toward Reflective Cooling. *Small* **16**, (2020).

14. Das Gupta, T. *et al.* Self-assembly of nanostructured glass metasurfaces via templated fluid instabilities. *Nat Nanotechnol* **14**, 320–327 (2019).

15. Badloe, T., Lee, S. & Rho, J. Computation at the speed of light: metamaterials for all-optical calculations and neural networks. *Advanced Photonics* vol. 4 Preprint at https://doi.org/10.1117/1.AP.4.6.064002 (2022).

16. Thureja, P. *et al.* Toward a universal metasurface for optical imaging, communication, and computation. *Nanophotonics* **11**, 3745–3768 (2022).

17. Wu, C. *et al.* Programmable phase-change metasurfaces on waveguides for multimode photonic convolutional neural network. *Nat Commun* **12**, (2021).

18. Wang, Z. *et al.* Metasurface on integrated photonic platform: From mode converters to machine learning. *Nanophotonics* vol. 11 3531–3546 Preprint at https://doi.org/10.1515/nanoph-2022-0294 (2022).

19. Wang, Q., Han, W., Wang, Y., Lu, M. & Dong, L. Tape nanolithography: a rapid and simple method for fabricating flexible, wearable nanophotonic devices. *Microsyst Nanoeng* **4**, (2018).

20.  Huang, W. *et al*. A mechanically bendable and conformally attachable polymer membrane microlaser array enabled by digital interference lithography. *Nanoscale* **12**, 6736–6743 (2020).

21.  Lorach, H. *et al*. Photovoltaic restoration of sight with high visual acuity. *Nat Med* **21**, 476–482 (2015).

22.  Sahu, R. R. *et al*. Single-step fabrication of liquid gallium nanoparticles via capillary interaction for dynamic structural colours. *Nat Nanotechnol* (2024) doi:10.1038/s41565-024-01625-1.

23.  Yin, Z. *et al*. Single-layer MoS 2 phototransistors. *ACS Nano* **6**, 74–80 (2012).

24.  Pradhan, N. R. *et al*. High Photoresponsivity and Short Photoresponse Times in Few-Layered WSe2 Transistors. *ACS Appl Mater Interfaces* **7**, 12080–12088 (2015).

25.  Hwang, I., Kim, J. S., Cho, S. H., Jeong, B. & Park, C. Flexible Vertical p-n Diode Photodetectors with Thin N-type MoSe2 Films Solution-Processed on Water Surfaces. *ACS Appl Mater Interfaces* **10**, 34543–34552 (2018).

26.  Ma, D. *et al*. Ultrathin GeSe Nanosheets: From Systematic Synthesis to Studies of Carrier Dynamics and Applications for a High-Performance UV-Vis Photodetector. *ACS Appl Mater Interfaces* **11**, 4278–4287 (2019).

27.  Watts, B. R. *et al*. Fabrication and performance of a photonic-microfluidic integrated device. *Micromachines (Basel)* **3**, 62–77 (2012).

28.  Chehadi, Z. *et al*. Soft Nano-Imprint Lithography of Rare-Earth-Doped Light-Emitting Photonic Metasurface. *Adv Opt Mater* **10**, (2022).

29.  Bar-On, O. *et al*. High Quality 3D Photonics using Nano Imprint Lithography of Fast Sol-gel Materials. *Sci Rep* **8**, (2018).

30.  Hou, H., Yin, J. & Jiang, X. Smart Patterned Surface with Dynamic Wrinkles. *Acc Chem Res* **52**, 1025–1035 (2019).

31.  Huang, Z., Hong, W. & Suo, Z. Evolution of wrinkles in hard films on soft substrates. *Phys Rev E Stat Phys Plasmas Fluids Relat Interdiscip Topics* **70**, 4 (2004).

32.  N Bowden wrinkles.

33.  Tomba, C. *et al*. Laser-Assisted Strain Engineering of Thin Elastomer Films to Form Variable Wavy Substrates for Cell Culture. *Small* **15**, (2019).

34.  Martinez, P. *et al*. Laser Generation of Sub-Micrometer Wrinkles in a Chalcogenide Glass Film as Physical Unclonable Functions. *Advanced Materials* **32**, (2020).

35.  Guo, C. F. *et al*. Path-guided wrinkling of nanoscale metal films. *Advanced Materials* **24**, 3010–3014 (2012).

36.  Zhou, L. *et al*. Regulating surface wrinkles using light. *Natl Sci Rev* **7**, 1247–1257 (2020).

37.  Liu, N. *et al*. Wrinkled Interfaces: Taking Advantage of Anisotropic Wrinkling to Periodically Pattern Polymer Surfaces. *Advanced Science* vol. 10 Preprint at https://doi.org/10.1002/advs.202207210 (2023).

38. Kim, H. S. & Crosby, A. J. Solvent-responsive surface via wrinkling instability. *Advanced Materials* **23**, 4188–4192 (2011).

39. Yang, Y., Duan, S. & Zhao, H. Highly Conductive Silicone Elastomers via Environment-Friendly Swelling and In Situ Synthesis of Silver Nanoparticles. *Adv Mater Interfaces* **8**, (2021).

40. Gallardo, A. *et al.* Chemical and Topographical Modification of Polycarbonate Surfaces through Diffusion/Photocuring Processes of Hydrogel Precursors Based on Vinylpyrrolidone. *Langmuir* **33**, 1614–1622 (2017).

41. Zeng, S. *et al.* Moisture-Responsive Wrinkling Surfaces with Tunable Dynamics. *Advanced Materials* **29**, (2017).

42. Harrison, C., Stafford, C. M., Zhang, W. & Karim, A. Sinusoidal phase grating created by a tunably buckled surface. *Appl Phys Lett* **85**, 4016–4018 (2004).

43. Tan, A., Ahmad, Z., Vukusic, P. & Cabral, J. T. Multifaceted Structurally Coloured Materials: Diffraction and Total Internal Reflection (TIR) from Nanoscale Surface Wrinkling. *Molecules* **28**, (2023).

44. Park, J. Y. *et al.* Controlled wavelength reduction in surface wrinkling of poly(dimethylsiloxane). *Soft Matter* **6**, 677–684 (2010).

45. Lu, H. *et al.* Controlled evolution of surface patterns for ZnO coated on stretched PMMA upon thermal and solvent treatments. *Compos B Eng* **132**, 1–9 (2018).

46. Stafford, C. M. *et al.* A buckling-based metrology for measuring the elastic moduli of polymeric thin films. *Nat Mater* **3**, 545–550 (2004).

47. Sun, J. Y., Xia, S., Moon, M. W., Oh, K. H. & Kim, K. S. Folding wrinkles of a thin stiff layer on a soft substrate. in *Proceedings of the Royal Society A: Mathematical, Physical and Engineering Sciences* vol. 468 932–953 (Royal Society, 2012).

48. Li, J., An, Y., Huang, R., Jiang, H. & Xie, T. Unique aspects of a shape memory polymer as the substrate for surface wrinkling. *ACS Appl Mater Interfaces* **4**, 598–603 (2012).

49. Li, Z. *et al.* Harnessing Surface Wrinkling–Cracking Patterns for Tunable Optical Transmittance. *Adv Opt Mater* **5**, (2017).

50. Chen, Z., Young Kim, Y. & Krishnaswamy, S. Anisotropic wrinkle formation on shape memory polymer substrates. *J Appl Phys* **112**, (2012).

51. Qi, L. *et al.* Writing Wrinkles on Poly(dimethylsiloxane) (PDMS) by Surface Oxidation with a CO2 Laser Engraver. *ACS Appl Mater Interfaces* **10**, 4295–4304 (2018).

52. Gao, N., Zhang, X., Liao, S., Jia, H. & Wang, Y. Polymer Swelling Induced Conductive Wrinkles for an Ultrasensitive Pressure Sensor. *ACS Macro Lett* **5**, 823–827 (2016).

53. Yang, Y. & Zhao, H. Water-induced polymer swelling and its application in soft electronics. *Appl Surf Sci* **577**, (2022).

54. Takahashi, M. *et al.* Photoinduced formation of wrinkled microstructures with long-range order in thin oxide films. *Advanced Materials* **19**, 4343–4346 (2007).

55. Lee, G., Zarei, M., Wei, Q., Zhu, Y. & Lee, S. G. Surface Wrinkling for Flexible and Stretchable Sensors. *Small* vol. 18 Preprint at https://doi.org/10.1002/smll.202203491 (2022).

56. Chung, J. Y., Nolte, A. J. & Stafford, C. M. Surface wrinkling: A versatile platform for measuring thin-film properties. *Advanced Materials* vol. 23 349–368 Preprint at https://doi.org/10.1002/adma.201001759 (2011).

57. Zhu, J. C., Zhou, J. K. & Shen, W. M. Polarisation-independent diffraction grating based on dielectric metasurface. *Electron Lett* **55**, 756–759 (2019).

58. Ma, T. *et al.* Dynamic Surface Wrinkles for in Situ Light-Driven Dynamic Gratings. *ACS Appl Mater Interfaces* **14**, 16949–16957 (2022).

59. Feng, J., Liu, Y. F., Bi, Y. G. & Sun, H. B. Light manipulation in organic light-emitting devices by integrating micro/nano patterns. *Laser and Photonics Reviews* vol. 11 Preprint at https://doi.org/10.1002/lpor.201600145 (2017).

## 9. Acknowledgements

The authors acknowledge the CRG grant (CRG/2022/004406) for funding the project. The authors also acknowledge support from CeNSE IISc facilities (Micro and Nano Characterization Facility (MNCF) and National Nanofabrication Center (NNFC)) funded by MHRD, MeitY and DST Nano Mission.

## 10. Author Contributions

TDG proposed the research direction and supervised the project. SB was involved in the fabrication, characterization, and mathematical modelling of the wrinkle formation mechanism, their corresponding simulations and data analysis. RRS was involved in the experimental setup for the mechanochromic sensor and in doing the post-process analysis of the corresponding data. RRS also did FIB of the samples. ODNS was involved in thin film metal depositions. SM helped SB in experimental setups. R did the coding for the diffraction spot simulations. MV made a PDMS heater for the Joule heating experiment. SB and TDG wrote the manuscript. All authors gave final consent to the manuscript.

## 11. Methods

### 11.1. Fabrication of Samples

*Preparation of PDMS and deposition of thin films*

Polydimethylsiloxane (PDMS) soft substrates are prepared by mixing the liquid PDMS base and cross-linking agent in various ratios (Dow Corning, Sylgard 184). The following notation PDMSx represents 1 part of curing agent is added to x parts of liquid PDMS base in weight ratio. Substrate softness increases with an increase in the proportion of liquid PDMS base and also with decrease in curing time. To fabricate wrinkles, PDMS 3, PDMS 5 and PDMS 10 are chosen as substrates. The mixture is stirred thoroughly and desiccated to remove the bubbles. The solution is poured onto a Si master with grooves and cured at 80°C in an oven for 3 hours and then peeled off the Si mould to obtain a cured soft substrate. Powdered $As_2Se_3$ is then thermally evaporated (HHV thermal evaporator) and deposited onto these substrates to form a thin film. Ag and Al thin films are deposited through DC magnetron sputtering (HHV sputtering tool). PMMA thin film is prepared by spin coating (EZspin-Apex Instruments spin coater) PMMA solution on PDMS substrate.

*Preparation of patterned Si molds*

The Si molds are patterned through photolithography using AZ4562 photoresist and etched through Reactive Ion Etching (RIE) using a RIE tool from Oxford instruments.

*Mechanical stretching of wrinkled thin films on PDMS*

The bilayer system is cut into a 6 cm × 1.5 cm rectangular shape. Both the sample ends are clamped onto stretching equipment operated through a Stepper motor controlled by Arduino. It is programmed for linear stretching

*Joule heating of wrinkled thin films on PDMS*:

A PDMS heater is prepared by encapsulating a zigzag network of nichrome wire with PDMS. The two ends of the nichrome wire are connected to a power supply (SCIENTIFIC, 30V-3A Dual Power Supply, PSD3203) using crocodile clips. The bilayer samples are kept over the PDMS encapsulation. The current passing through the nichrome wires is controlled through the source meter.

White light is made normally incident on the sample, and the spectrometer detector is placed at an angle of 35° with respect to the sample surface.

*Simulation for determining outcoupling efficiency of encapsulated light sources:*

Finite difference time domain algorithm in the commercially available software Lumerical is used for simulations. The wavelength of the wrinkles is determined from SEM images and amplitude from AFM data. The PDMS substrate is a homogeneous material with a refractive index of 1.4. The optical constants for PMMA are imported as the real and imaginary parts of the refractive index. The finite-difference time-domain (FDTD) region is chosen as 50μm × 50 μm, with periodic boundary conditions along the x and y directions and PML boundary condition along z direction. The area chosen is large enough to encompass 5-6 wrinkles.

A plane wave source (wavelength range 400nm-700nm) is placed within the PDMS substrate to imitate an encapsulated white light source. The frequency-domain field and power monitor are suspended in air,10µm above the wrinkled/non-wrinkled bilayer system to obtain the power of the transmitted light

*Simulation for diffraction spots*:

The simulation of the diffraction spots is done through MATLAB coding. The periodicity and amplitude of the wrinkles is determined from AFM images. Detailed simulation process is illustrated in Supplementary Information section 3.6.

## 11.2. Thermal Processing

The samples are heated to different temperatures within the range of 80°C to 120°C in intervals of 10°C followed by immediate cooling to 25°C room temperature to get wrinkles of desired periodicity and orientation.

## 11.3. Material Characterization

*Scanning Electron Microscopy (SEM):* The scanning electron microscope images are taken with Zeiss-Gemini SEM, ULTRA 500 model in the secondary electron (SE2) mode. The nonconducting samples ($As_2Se_3$, PMMA thin films) are sputtered with 5nm gold to form a conducting layer for electronic charge dissipation.

*Atomic Force Microscopy (AFM):* Park AFM NX20 tool and TESPA V2 AFM tip is used to obtain the surface morphology of the wrinkles (periodicity and amplitude) in the non-contact mode.

*Focused ion beam (FIB):* Thermo Fisher Scientific Scios Dual Beam focused ion beam (FIB)–SEM system is used to make TEM lamella of Ga-deposited PDMS, which were placed on a copper grid for cross-section imaging in secondary electron mode.

## 11.4. Optical Characterization

*The Ocean Optics FLAME series UV–visible spectrophotometer*: used to obtain the reflectivity at normal incidence, with the spectrophotometer detector placed at an angle of 35° with respect to the sample surface. The package python-Seabreeze is used to access the spectrophotometer in a Python program. The reflectivity thus obtained is used for real-time computation of the sample's CIE x and y coordinates and hue. The packages numpy and matplotlib are used for array manipulation and visualization (graphs and animations) respectively. The reflectivity spectra obtained are smoothened by box averaging around a small window of the raw data points to remove device-sourced

*Power-meter*: Gentec EO power-meter is used for registering the output power of the encapsulated LEDs.

*LASER Pointers*: Handheld LASER pointers, fixed on a stable LASER holder, are used for diffraction experiments.

## Competing interests

Th authors declare no conflicting interests.

# Supplementary Information

Tailored Thin Films: Modulating Soft Photonics with Dynamically Tunable Large Area Microstructures via Controlled Thermal Processing

Srijeeta Biswas, Renu Raman Sahu, Omkar Deokinandan Nayak Shinkre , Shubham Meena, Ramnishanth ,Mark Vailshery, Tapajyoti Das Gupta[*]

Laboratory of Advanced nanostructures for Photonics and Electronics, Department of Instrumentation and Applied physics, Indian Institute of Science, C.V. Raman Road, Bengaluru-560012, India

E-mail: tapajyoti@iisc.ac.in

TABLE OF CONTENTS

# 1. METHODS

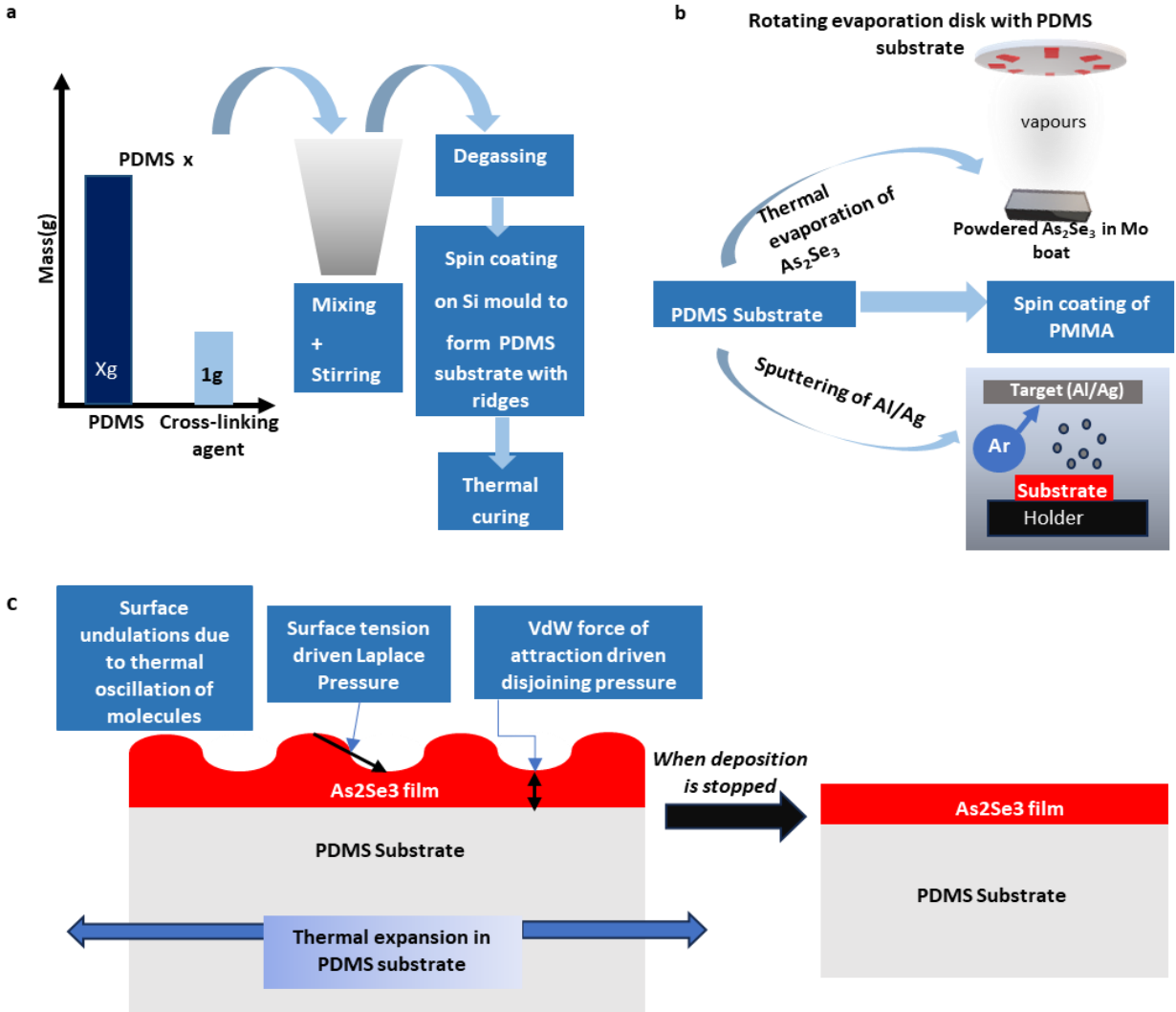

Figure 1| **Schematics for sample preparation and thin film formation.** a, Fabrication of PDMS substrates of different ratios. b, Various methods of deposition of thin films on a PDMS substrate. c, Mechanism of the formation of thin uniform films on a substrate

## 1.1 Preparation of PDMS

Soft and flexible Polydimethylsiloxane (PDMS) substrates are prepared by mixing X part of liquid PDMS base (Sylgard 184, Dow Corning) with 1 part of curing agent in weight ratio to form PDMSx. Substrate softness increases with an increase in the proportion of the liquid PDMS base i.e. PDMS 10 is softer than PDMS 5 which is again softer than PDMS 3 and so on. The mixture is stirred thoroughly and desiccated to remove all bubbles. The solution is then poured onto Si moulds (containing grooves pre-etched through photolithography and RIE-F) and cured at 80°C for 3 hours. Then the cured PDMS substrate (containing ridges) is peeled off from the Si mould (SI Fig.1a).

## 1.2 Selection of thin film and substrate materials

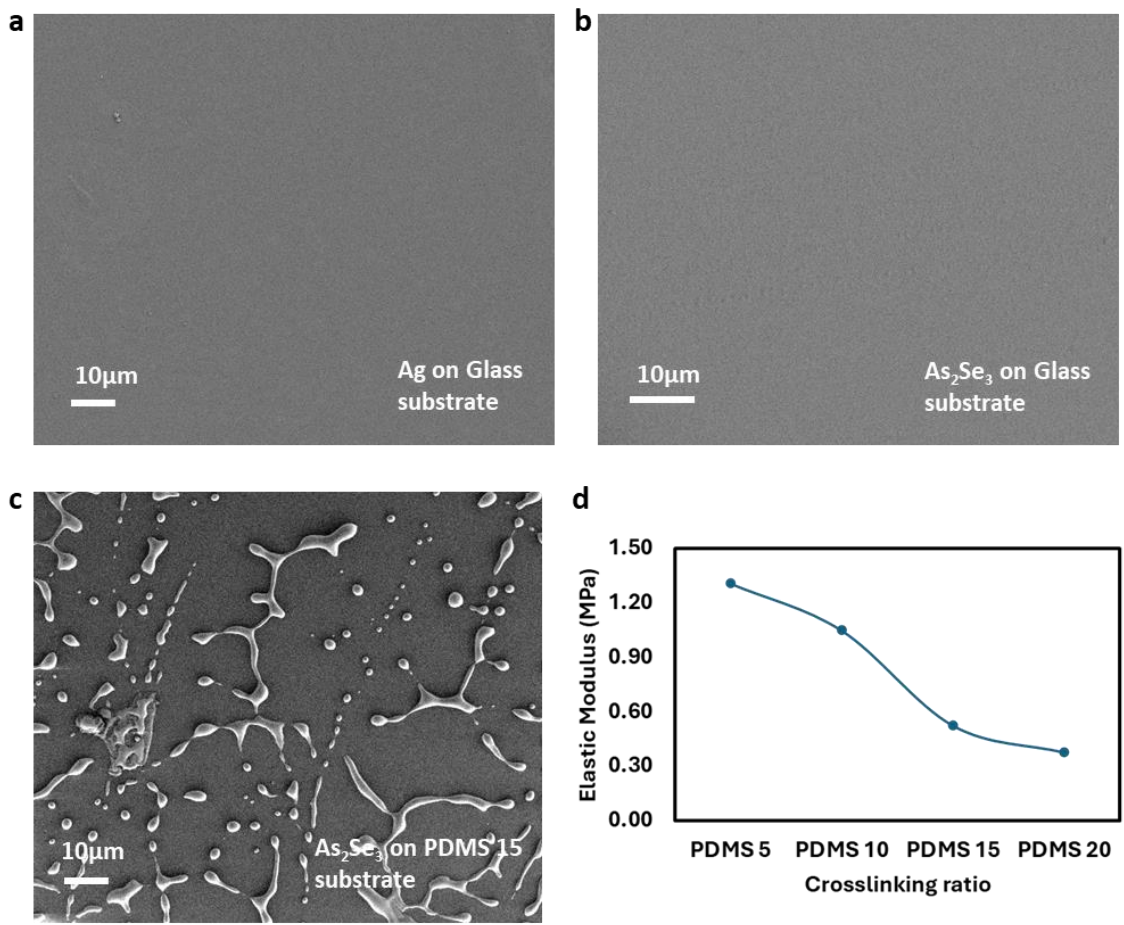

*Figure 2| a-b, Absence of wrinkles on rigid substrate (glass). c, Absence of wrinkles on very soft substrate (PDMS 15) due to dewetting. d, Elastic modulus vs. cross-linking ratio of PDMS substrates cured at 80°C for 3 hours.*

To achieve well-ordered wrinkles, the substrate's thermal expansion coefficient needs to be carefully balanced. It shouldn't expand excessively when heated, risking cracks in the thin film above as seen with PDMS 15 and PDMS 20 as substrates. Young's modulus being inversely proportional to the thermal expansion coefficient, the thermal expansion coefficient of PDMS 5 and 15 can be calculated through simple ratio proportion method. The thermal expansion coefficient of PDMS 10 is $\approx 270 \times 10^{-6}/K$ and from SI Fig. 2d we see that its Young's modulus is $1.05\ MPa$. The thermal expansion coefficient of PDMS 15, having a Young's modulus of $0.53\ MPa$, is $\approx 535 \times 10^{-6}/K$ and that of PDMS 5, having a Young's modulus of $1.31\ MPa$, is $\approx 216 \times 10^{-6}/K$. The substrate also shouldn't contract too less during cooling, failing to provide enough stress for the thin film atop it to form uniform wrinkles, as seen with solid substrates like glass and silicon. Thus, PDMS 3, PDMS 5 and PDMS 10 are found to be the best suited for our process, their thermal expansion coefficient being in the range $200 - 350 \times 10^{-6}/K$ [1]

The thin film's thermal expansion coefficient ought to be significantly lower than that of the substrate. This ensures that during cooling, the film contracts much less than the substrate, subjecting it to substantial compressive stress, surpassing its critical stress threshold, induced by the substrate.

| Thin film material | Thermal expansion coefficient | Surface energy | Young's modulus |
|---|---|---|---|
| As₂Se₃ | $20 - 90 \times 10^{-6}/K$ [2-4] | $0.2 J/m^2$ [5] | $1 \times 10^{10} Pa$ [4] |
| Al | $18 - 30 \times 10^{-6}/K$ [6] | $0.06 - 0.07 J/m^2$ [7] | $4 - 7 \times 10^{10} Pa$ [8] |
| Ag | $19 - 31 \times 10^{-6}/K$ [9] | $0.045 - 0.059 J/m^2$ [10] | $3 - 8 \times 10^{10} Pa$ [11] |
| PMMA | $40 - 90 \times 10^{-6}/K$ [12] | $0.0377 - 0.0588 J/m^2$ [13] | $3 - 4 \times 10^{9} Pa$ |

*Table 1| Thermal expansion coefficient, surface energy and Young's modulus of the thin film materials used.*

The wavelength of the wrinkles formed is inversely proportional to the surface energy of the thin film. As₂Se₃ having much higher surface energy than that of Al and Ag, forms wrinkles of much smaller wavelength than the others. The Al and Ag films easily form oxides and turn yellowish when exposed to ambient atmosphere. This oxide formation is reduced significantly by curing the PDMS substrates at 80°C for more than 24 hours.

Moreover, the melting point or the glass transition temperature of the substrates and the thin films must be greater than our maximum working temperature ($i.e. > \approx 200°C$).

As₂Se₃ has generated the most well-organized 1D and 2D sinusoidal wrinkles.

### 1.3 Preparation of As₂Se₃ (chalcogenide) thin film

Powdered As₂Se₃ is thermally evaporated (HHV thermal evaporator) and deposited onto the PDMS substrates to form thin films. The thickness of the film and the temperature of the substrate are monitored by the respective inbuilt thickness and temperature monitoring sensors. The films are deposited at a rate of 2.5 Å/s. The substrates are kept at room temperature (SI Fig. 1b).

## 1.4 Preparation of Ag and Al (metals) thin films

The deposition of the metallic thin films is done by using direct current (DC) magnetron sputtering deposition technique (HHV sputtering tool). The film thickness deposited in a given time is measured through optical profilometry and the sputtering tool is accordingly calibrated to give the required film thickness. The depositions are carried out at room temperature (SI Fig. 1b). Al depositions are done at $6 \times 10^{-3}$ Argon pressure and 16W sputtering power. Ag depositions are done at $9 \times 10^{-3}$ Argon pressure and 16W sputtering power.

## 1.5 Requirements and explanation for the formation of uniform thin films

When the film deposition initiates, surface undulations occur in the film due to the thermal oscillation of the molecules. The Vander Waal's force of attraction driven disjoining pressure favours these undulations whereas the surface tension driven Laplace pressure tries to homogenize the film. Beyond a critical film thickness (which varies from material to material depending on its Vander Waal's force of attraction), the Laplace pressure predominates over the disjoining pressure and a uniform thin film is formed on the substrate. Below this critical thickness uniform film depositions do not occur. This critical thickness for As$_2$Se$_3$ is considerably higher ($\approx 60 - 70 nm$) as compared to materials like Al or Ag, where uniform thin films can be achieved at much lower thicknesses ($\approx 10 - 15 nm$) (SI Fig. 1c).

During the deposition the substrate is constantly struck by the hot molecules of the thin film due to which it expands. If the substrate is soft enough (i.e. having a high thermal expansion coefficient) it may form cracks, thereby cracking the thin film deposited over it. The higher the time of deposition more the chances of crack formation. To prevent this, the substrate should have an optimum softness, and the rate of deposition must be high enough so that the duration of deposition decreases.

## 1.6 Preparation of PMMA (polymer) thin film

495PMMA A2 resist is spin coated on the PDMS substrate at 4000 rpm for 40 seconds followed by hard baking at $110°C$ for 2 minutes to form a thin film on the substrate.

## 2. MATHEMATICAL MODELING OF WRINKLED THIN FILM FORMATION ON PDMS SUBSTRATES

In this section we hypothesize the wrinkle forming mechanism in rigid thin films deposited atop soft flexible substrates. Also, we propose a Wavelength Equation, and a Critical Length Equation based on a few assumptions to model the attributes of the experimentally observed wrinkles.

### 2.1 Thermal processing and explanation of stress formation

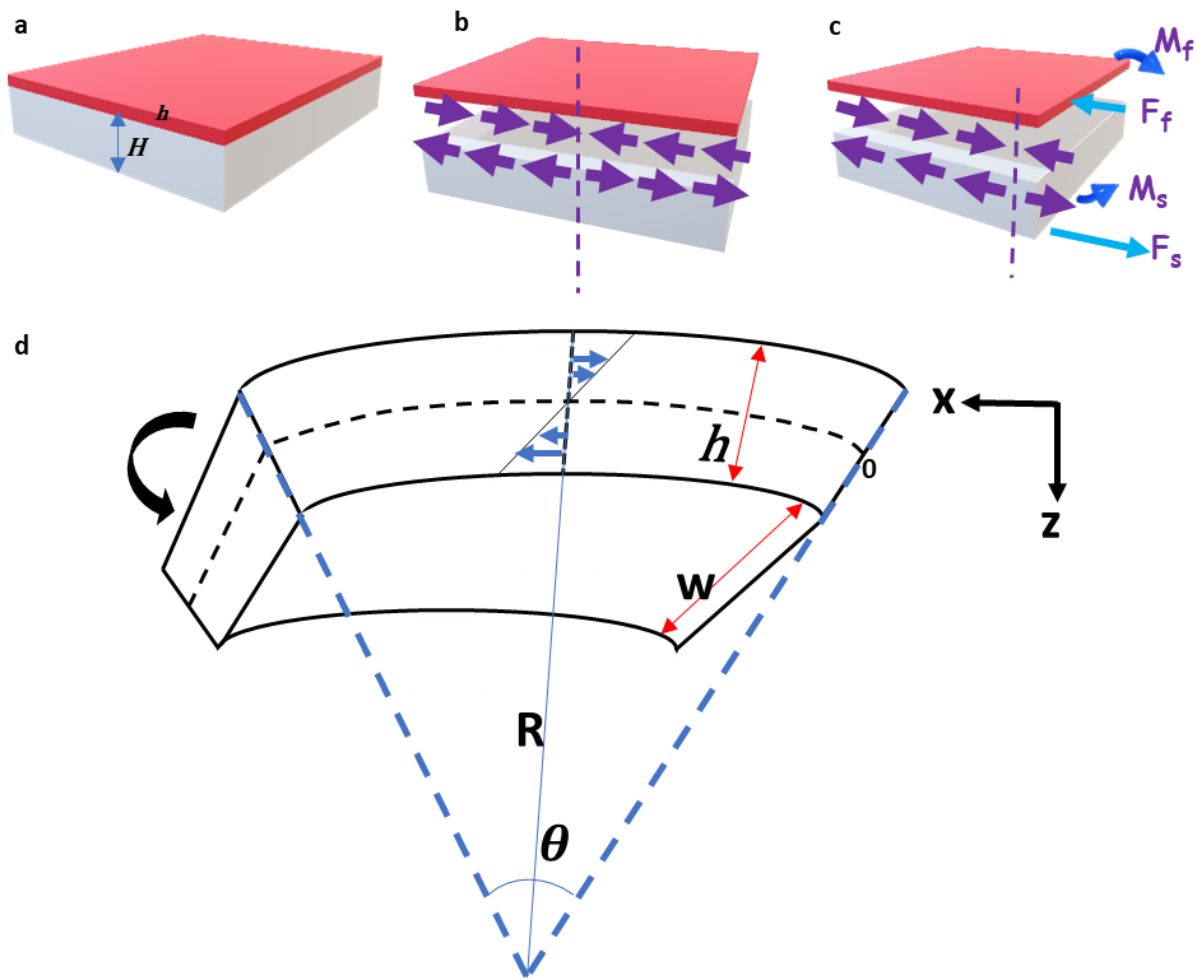

*Figure 3|**Stress formation mechanism in a film** a, Schematic of a film-substrate bilayer system with film thickness h and substrate thickness H. b, Schematic of the stress formation in the film and the substrate during cooling. Purple arrows indicate the direction of stress. c, Schematic representing the resultant stress and moment acting on the film and the substrate. d, Schematic showing the stress formation on the film of width w and radius of curvature R, due to the compressive stress acting over it*

When the film-substrate bilayer system as shown in SI Fig.3a starts cooling down, the PDMS substrate, owing to its higher thermal expansion coefficient than that of the film deposited atop, contracts faster than the film. In order to maintain the compatibility between the film and the substrate, the substrate experiences a tensile stress whereas the film experiences a compressive stress (depicted in SI Fig. 3b). In SI Fig. 3c we have replaced the interfacial set of stresses by a single force and moment for each of the film and the substrate, $F_f$ and $M_f$ for the film and $F_s$ and $M_s$ for the substrate. The system will bend to counteract the unbalanced moments. Considering the film to be a longitudinal beam and defining the strain over it as, (SI Fig. 3d)

$$Strain = \frac{\{R\theta - (R-h)\theta\}}{R\theta} \tag{1}$$

Therefore, the stress acting over it is defined as,

$$\sigma_f = \frac{Y_f\{R\theta - (R-h)\theta\}}{R\theta}$$

$$\Rightarrow \sigma_f = \frac{Y_f h}{R} \tag{2}$$

Where $\sigma_f$ is the stress acting on the film, $Y_f$ is its Young's modulus, $R$ is the radius of curvature, $\theta$ is the angle of curvature and h is the film thickness. Thus, we can see that for a given film stiffness and radius of curvature, higher the film thickness higher is the stress experienced by it. This is the reason why we get wrinkles of larger wavelengths for higher film thicknesses and wrinkle formation also becomes easier with increase in film thickness.

## 2.2 Hypothesis of wrinkle formation mechanism and our proposed Wavelength Equation

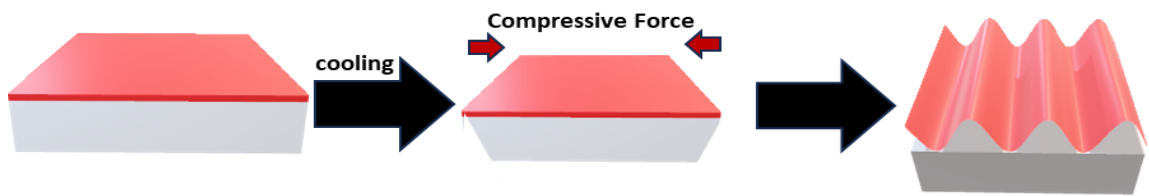

*Figure 4|* **Schematic showing the wrinkle formation mechanism in the thin film.** *Due to the mismatch in the thermal expansion coefficient of the film and the substrate, during cooling, wrinkles are formed in the film to maintain the equilibrium of the system.*

When the system is heated, the substrate expands more than the film, thereby exerting a tensile stress over the film. If the substrate expansion is too high, as in the case of PDMS 15,20 and above, the tensile stress acting on the film is so high that cracks are formed in it (SI Fig.2c).

As the system cools, the substrate contracts more than the film, resulting in a compressive stress on the film. To maintain equilibrium, wrinkles with appropriate wavelength form in the film, as shown in SI Fig.4 The size of these wrinkles directly correlates with the stress on the film. A greater mismatch in thermal expansion coefficients between the film and substrate leads to increased stress on the film, resulting in wrinkles with larger wavelength. Similarly, cooling the system from a higher temperature increases the stress on the film, again enlarging the wrinkles' wavelength.

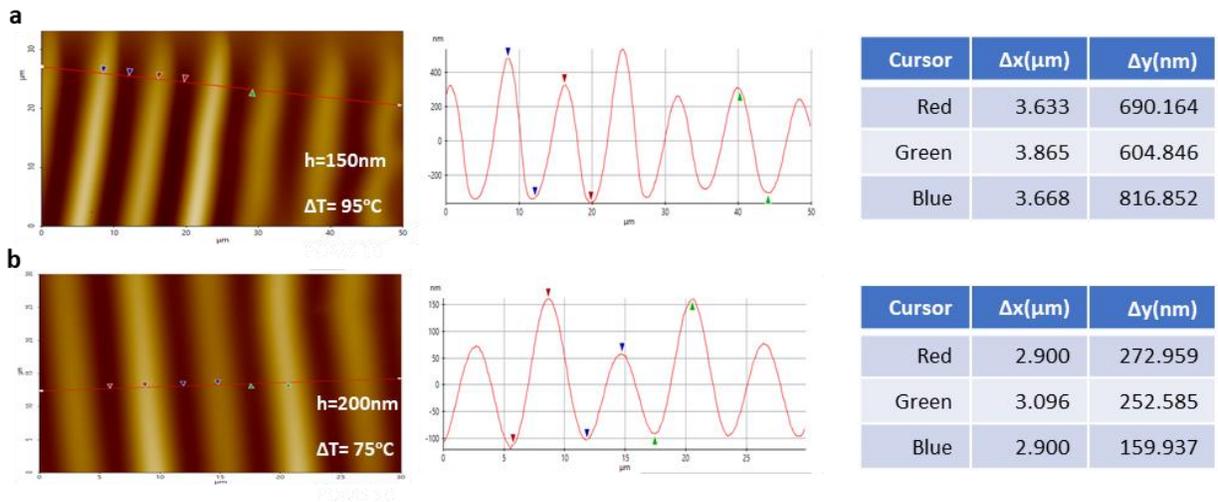

Figure 5 | **AFM images illustrating that the wavelength of the wrinkles is much greater than their amplitude.** Samples having a film of $As_2Se_3$ deposited over PDMS 5 substrate. The line plot shows the surface profile of the film across the line drawn over the AFM image. The table to the right gives the Δx(μm) and Δy(nm) values of the cursor pairs in the line profile. Δx(μm) is half of the wavelength of the wrinkles whereas Δy(nm) is their amplitude. a, for a film thickness of 150nm, and change is temperature of 95°C, the wavelength and amplitude of the wrinkles are around 7.5μm and 700nm respectively. b, for a film thickness of 200nm, and change is temperature of 75°C, the wavelength and amplitude of the wrinkles are around 6μm and 200nm respectively. Thus, the amplitude of the wrinkles is much smaller than their wavelengths.

**MATHEMATICAL MODELING**

Let us consider that stress energy $U$ leads to the formation of n number of wrinkles

$$U = \int_0^U dU$$

$$= \int_0^l f dx$$

Where $f$ is the compressive force acting on the film at an instant and $dx$ is the contraction in the film along the direction of the compressive force at that instant. $l$ is the total contraction that should occur in the film due to the total stress energy $U$.

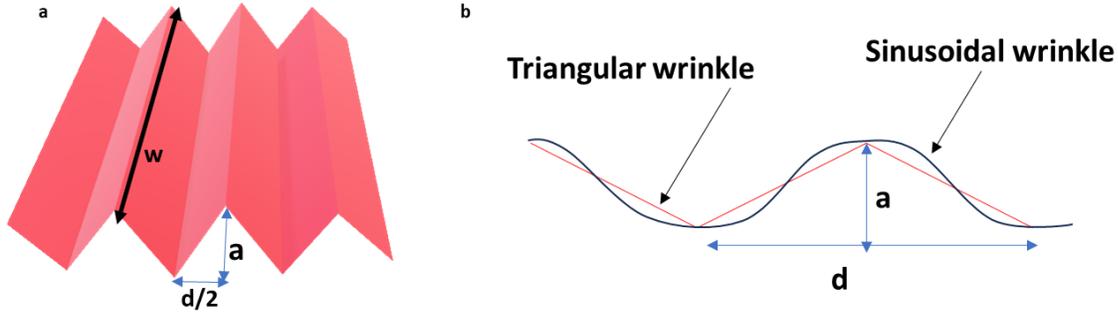

*Figure 6| **Schematic of an array of wrinkles**. a, Width w, amplitude a, and periodicity (wavelength) d. For modeling purpose, we have assumed the wrinkles to be triangular. b, Schematic representing the superposition of sinusoidal and triangular wrinkles of same amplitude and wavelength. Sinusoidal wrinkles have higher surface are than triangular ones.*

The Young's modulus of a system is defined as

$$Y = \frac{\sigma}{\epsilon}$$

Where $Y$ is the Young's modulus of the system, $\sigma$ is the stress and $\epsilon$ is the strain acting on it. The strain acting on the film in the bilayer system can be defined as

$$\epsilon = (\alpha_s - \alpha_f)Y_f$$

Where $Y_f$ is the Young's modulus of the film, $\alpha_s$ and $\alpha_f$ are the thermal expansion coefficients of the substrate and the film, respectively.

Thus, the compressive force $f$ acting on the film at an instant is defined as

$$f = Y_f(\alpha_s - \alpha_f)\Delta T w h \qquad (3)$$

$\Delta T$ signifies the temperature change within system and $w, h$ are the width and thickness of the film, respectively. Therefore, the stress energy $U$ can be defined as

$$U = \int_0^l Y_f(\alpha_s - \alpha_f)\Delta T(wh)dx$$

$$= Y_f(\alpha_s - \alpha_f)\Delta T w h l \qquad (4)$$

Total surface area of a triangular wrinkle can be defined as (assuming triangular wrinkles as shown in SI Fig.6a, and using Pythagoras theorem,)

$$A = 2 \times 2w\sqrt{\frac{d^2}{4} + a^2}$$

Therefore, the surface area of a sinusoidal wrinkle will be defined as

$$A = C \times 2 \times 2w\sqrt{\frac{d^2}{4} + a^2} \tag{5}$$

Where $a$ is the amplitude and $d$ is the periodicity(wavelength) of the wrinkle. The first 2 is for the top and the bottom surface of the wrinkle and the second 2 is for both the faces of the triangle. Only the bottom surface of the wrinkle is the newly formed surface during wrinkle formation. $C$ is a proportionality constant, greater than 1, which accounts for the disparity in surface area between a sinusoidal and a triangular wrinkle. The surface area of a sinusoidal wrinkle is greater than that of a triangular wrinkle as shown in SI Fig.6b. Higher the value of the amplitude and periodicity of the wrinkles, higher is the value of $C$.

At the verge of wrinkle formation, the stress energy gets converted to the surface energy of the newly formed surfaces of the wrinkles. Thus, from energy conservation we get

$$Y_f(\alpha_s - \alpha_f)\Delta T whl = Cn2w\sqrt{\frac{d^2}{4} + a^2}\gamma_f \tag{6}$$

Where $\gamma_f$ is the surface energy of the thin film.

Now, since $d \gg a$ (refer SI Fig. 5), using Binomial expansion we get,

$$Y_f(\alpha_s - \alpha_f)\Delta T whl = C\gamma_f nwd \tag{7}$$

$$d = \frac{Y_f(\alpha_s - \alpha_f)\Delta T hl}{Cn\gamma_f} \tag{8}$$

We know that the strain acting on the film is defined as,

$$\in = \frac{l}{L} = (\alpha_s - \alpha_f)\Delta T \tag{9}$$

Where $l$ is the total change in length of the film, along the direction of the compressive stress and $L$ is the total expanded length of the film after heating, along which the compressive stress starts acting when the system starts cooling down from the maximum temperature. Therefore,

$$l = L(\alpha_s - \alpha_f)\Delta T \tag{10}$$

This leads to our proposed Wavelength Equation,

$$d = \frac{Y_f[(\alpha_s - \alpha_f)\Delta T]^2 hL}{Cn\gamma_f} \tag{11}$$

Ascertaining the maximum length $L$ of the film (i.e. the length of the film after heating) and the proportionality constant $C$ directly isn't feasible. Therefore, $(L/C n)$ has been replaced by a proportionality constant $C_w$. It is a dimensional constant with units of length (i.e. meter when SI units are used for all other parameters). Thus, our final wavelength equation is,

$$d = C_w \frac{Y_f \, [(\alpha_s - \alpha_f)\Delta T]^2 h}{\gamma_f} \qquad (12)$$

The proportionality constant $C_w$ depends on the following parameters

i. More the **thermal expansion coefficient of the substrate** more is the tensile stress exerted by it on the film during heating. $L$ is directly proportional to the tensile stress acting on the film. In case of PDMS, the thermal expansion coefficient varies with the curing temperature and also the duration of curing.
ii. More the **thermal expansion coefficient of the film** more is its expansion during heating, and more is the value of $L$. In case of thin films, the thermal expansion coefficient of a given material can also vary with the film thickness.
iii. Lower the **surface energy of a film,** the higher the surface area required to balance out a given stress. Thus, to balance out a given stress during heating, the expansion in length $L$ will be more for a thin film material having lower surface energy. Thus, $L$ is inversely proportional to surface energy.
iv. The higher the **tensile force acting on the film** while heating, higher is the expansion in it. Thus $L$ is directly proportional to the tensile force acting on the thin film. The tensile force is dependent on the **Young's modulus ($Y_f$), width ($w$), and thickness of the film ($h$), change in temperature of the system ($\Delta T$), and the thermal expansion coefficient mismatch between the film and the substrate ($\alpha_s - \alpha_f$)**. (Refer SI equation 3).
v. $L$ also depends on the initial length of the sample. Higher the initial length, higher is the expanded length $L$. However, it has been experimentally seen that beyond a critical length $L_c$, keeping other parameters constant, as $L$ increases, $n$ also increases such that the wrinkle periodicity doesn't change significantly. Below the critical length wrinkle formation does not take place.

## 2.3 Comparison of results from Wavelength Equation with experimental results

To model the above picture, we solve the Wavelength Equation for bilayer systems consisting of As$_2$Se$_3$ thin films deposited on PDMS substrates with arbitrary but reasonable choice of values for coefficients of thermal expansion for the substrate and film, and also for the surface energy and Young's modulus of the As$_2$Se$_3$ thin film. The selected values being $\alpha_{PDMS\,5} = 216 \times 10^{-6}/K$, $\alpha_{PDMS\,10} = 270 \times 10^{-6}/K$, $\gamma_{As_2Se_3} = 0.2 J/m^2$, $Y_{As_2Se_3} = 1 \times 10^{10} Pa$. The values lie within the respective ranges obtained from literature as enlisted in SI Section 1.2.

As shown in the following figures, for each trend in wrinkle wavelength periodicity as observed from the top view scanning electron microscope images, there exist parameters of Wavelength Equation which, when used to solve, give matching results.

### 2.3.1 Dependence of wrinkle wavelength on film thickness and temperature

Our Wavelength Equation predicts an increase in wrinkle wavelength with increase in film thickness and temperature.

| Film thickness(nm) | Proportionality constant ($C_w$) with PDMS 5 substrate (m) |
|---|---|
| 100 | $3.07 \times 10^{-6}$ |
| 150 | $3.11 \times 10^{-6}$ |
| 200 | $3.155 \times 10^{-6}$ |

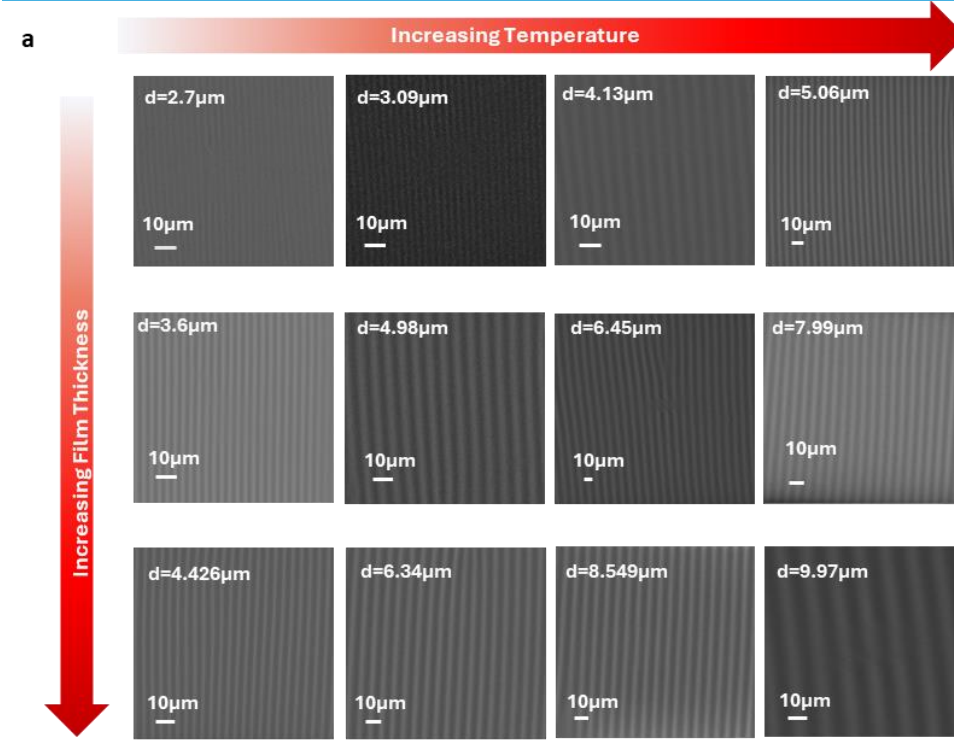

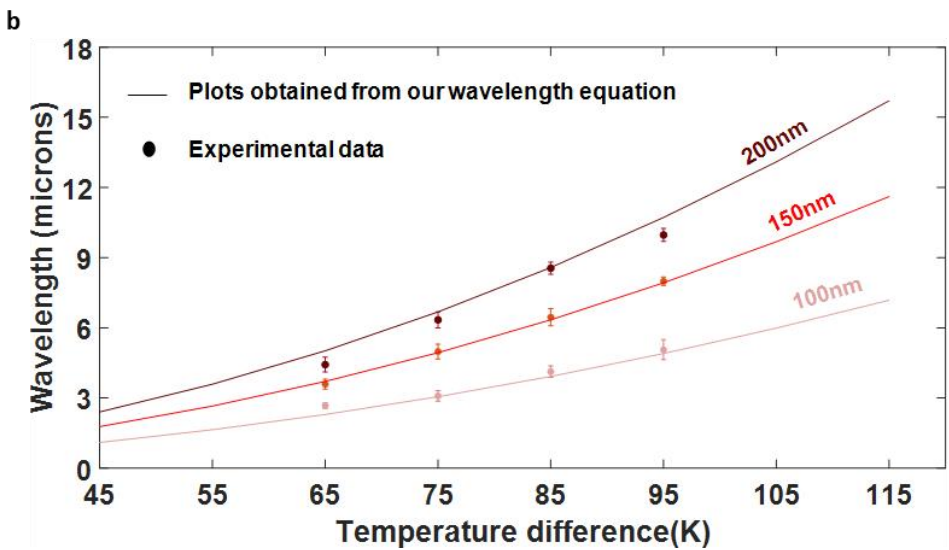

*Figure 8| (Top) Table showing the parameters used in the Wavelength Equation, for different film thicknesses, matching the modeling results with the experimental results. a, Comparison matrix of SEM images depicting the dependence of wrinkle wavelength on Film thickness and Heating Temperature. As the film thickness and heating temperature increases, the wavelength (d) also increases. Substrate: PDMS 5, Thin Film material: $As_2Se_3$. Heating temperature ranges from 90°C to 120°C, in intervals of 10°C and the room temperature is 25°C. The film thickness ranges from 100nm to 200nm in intervals of 50nm.*

According to our Wavelength Equation, an increase in film thickness and/or $\Delta T$ leads to an increase in the wavelength of the wrinkles. As the film thickness increases, the stress acting on it during the cooling process also increases, thereby increasing the wavelength (SI Equation 2). Similarly, as the heating temperature increases, the compression in the substrate while cooling down to room temperature also increases, thereby increasing the compressive stress on the film during the cooling process (SI Equation3).

Fig.8a depicts the dependence of wrinkle wavelength on film thickness and heating temperature. As the film thickness and heating temperature increases, the wavelength(d) also increases. The substrates are PDMS 5, with the thin film material being $As_2Se_3$. In Fig. 8a the heating temperature ranges from 90°C to 120°C, in intervals of 10°C and the room temperature (i.e. the cooling temperature) is 25°C. The film thickness ranges from 100nm to 200nm in intervals of 50nm. Fig. 8b illustrates the compatibility of our wavelength equation with the experimental results. The solid circles represent data from the experiments while the solid lines depict the corresponding plots obtained from the Wavelength Equation using similar parameters (i.e. thermal expansion coefficient of substrate and film, thickness, surface energy, Young's modulus of film and $\Delta T$).

### 2.3.2 Dependence of wrinkle wavelength on thermal expansion coefficient mismatch between film and substrate

Our Wavelength equation predicts that as the difference in the thermal expansion coefficient between the film and the substrate increases, the wrinkle wavelength also increases.

As the difference in thermal expansion coefficients between the substrate and the rigid film deposited on it increases, so does the disparity in their thermal expansion. Consequently, as the system cools, the mismatch in contraction between the substrate and the film intensifies, leading to greater compressive stress on the film. This is why we observe larger wrinkles when there's a higher thermal expansion coefficient mismatch between the film and the substrate. PDMS10 has a higher thermal expansion coefficient than PDMS5, hence the wrinkles formed on the thin film deposited atop PDMS10 have higher wavelength than that deposited on PDMS5, for the same film material and film thickness (SI Fig.10). It is to be noted that mismatch between the theoretical and experimental values arises because all the thin film and substrate parameters have been assumed to be constant with respect to change in film thickness and system temperature. Also, in the theoretical model the sinusoidal wrinkles have been assumed to be triangular.

| Film thickness(nm) | $C_w\ (m)$ with PDMS 5 as substrate | $C_w\ (m)$ with PDMS 10 as substrate |
|---|---|---|
| 100 | $3.07 \times 10^{-6}$ | $3.15 \times 10^{-6}$ |
| 200 | $3.155 \times 10^{-6}$ | $3.2 \times 10^{-6}$ |

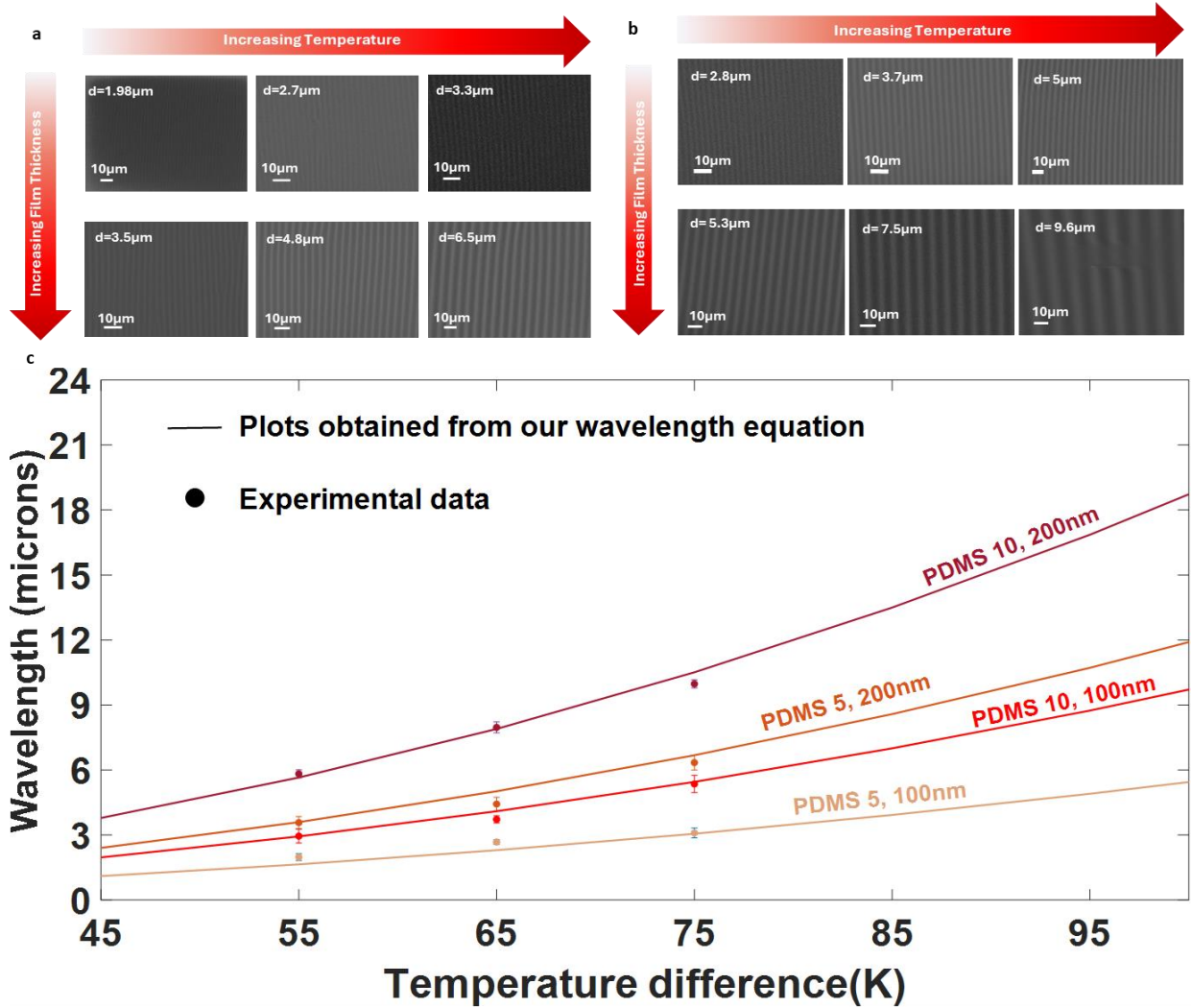

*Figure 9| (Top) Table showing the parameters used in the Wavelength Equation, for different film thicknesses for PDMS 5 and PDMS 10 substrates, matching the modeling results with the experimental results. a, Comparison matrix of SEM images with PDMS 5 as substrate. b, Comparison matrix of SEM images with PDMS 10 as substrate. (a-b), The temperature varies from 80°C to 100°C in intervals of 10°C . The film thickness are 100 nm and 200nm. Film material: As$_2$Se$_3$  c, Plot showing the dependence of wrinkle wavelength on the thermal expansion coefficient mismatch between the film and the substrate. The solid dots represent the experimental data points, and the solid lines are obtained from our Wavelength Equation using similar parameters.*

### 2.3.2.1 Role of PDMS curing time on the thermal expansion coefficient mismatch between the film and the substrate

As the curing time of PDMS increases, the crosslinking between the PDMS base and the curing agent also increases, thereby reducing its softness i.e. with increase in curing time the thermal expansion coefficient of PDMS decreases

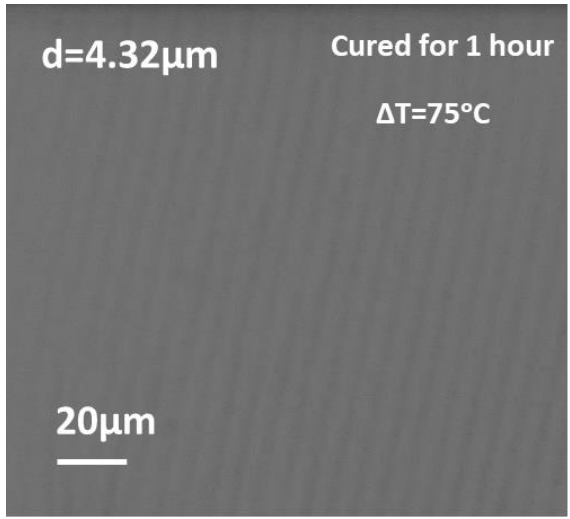
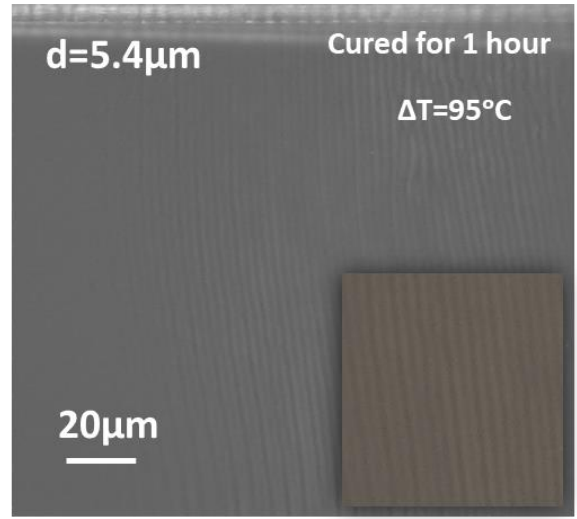
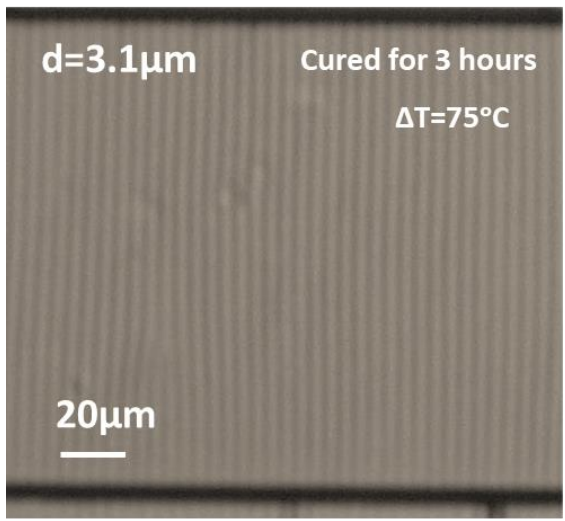
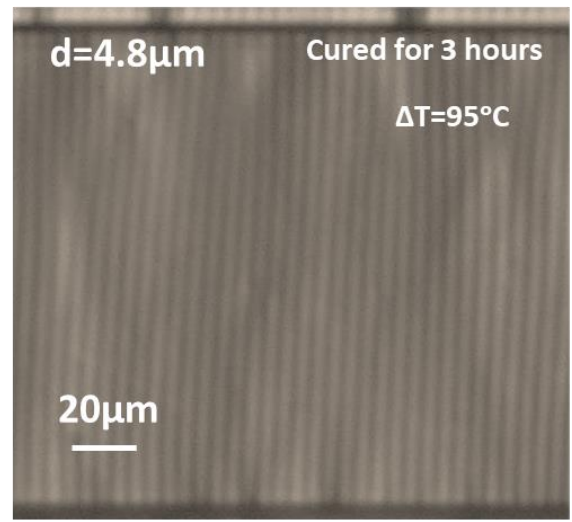
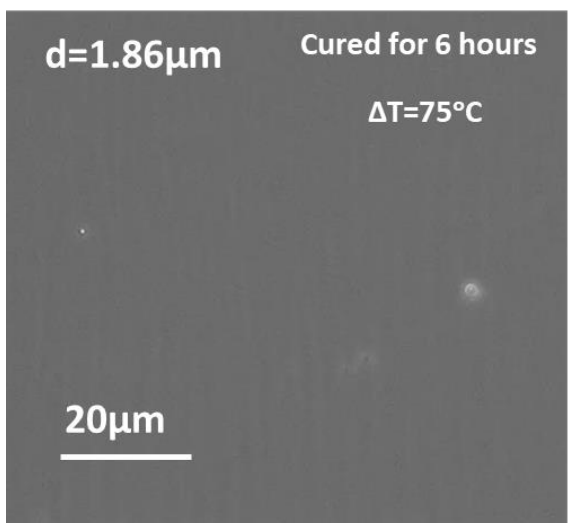
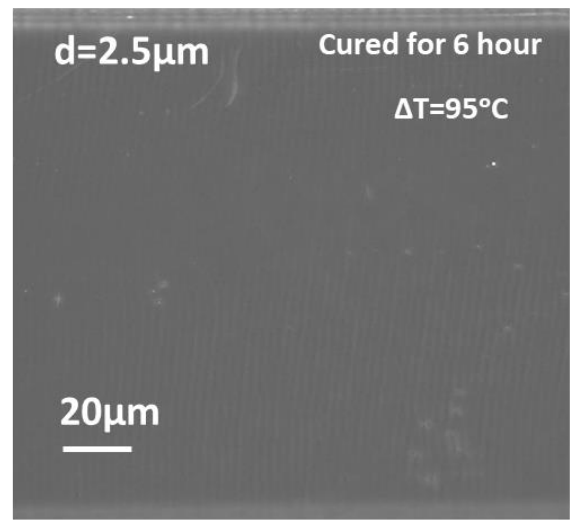

*Figure 10| This figure illustrates that, for the same film thickness, as the curing time of PDMS increases, the wavelength of the wrinkles formed in the bilayer system decreases. Substrate: PDMS 5, Thin film material: $As_2Se_3$.*

As the thermal expansion coefficient of the substrate decreases with increase in curing time, the mismatch in the thermal expansion coefficient between the film and the substrate also decreases, thereby reducing the stress acting on the film. Thus, by increasing the PDMS 5 curing time from 1 hour to 3 hours, the wavelength of wrinkles decreases from $4.32 \mu m$ to $1.86 \mu m$ for $\Delta T = 75°C$ and from $5.4 \mu m$ to $2.5 \mu m$ for $\Delta T = 95°C$. All the samples consist of 100nm $As_2Se_3$ deposited on PDMS 5 substrates.

When the compressive stress acting on the thin film in a single direction is high enough, non-linearities take place leading to paired wrinkles (Fig. 10 Top right). Such high stresses lead to high aspect ratio wrinkles (i.e. wrinkle amplitude being much higher than their wavelength), which merge at the wrinkle tips to form doubly paired wrinkles as illustrated in Fig. 10 Top right inset. Each doubly paired wrinkle has a very shallow indent at the top and acts as a single wrinkle.

### 2.3.3 Requirement of a critical film thickness for wrinkle formation

Compressive stress acting on a thin film is directly proportional to the thickness and Young's modulus of the film (Equation 3). Wrinkles manifest in a film when the stress energy due to a compressive force is balanced by the surface energy generated due to the formation of new surfaces. Consequently, a thin film with a higher surface energy will require a higher compressive stress for wrinkle formation. For instance, when being cooled down from 80°C to 25°C, $As_2Se_3$, having a Young's modulus of $10^{10} Pa$, and surface energy of $0.2 \ J/m^2$ requires a critical thickness exceeding 70nm for wrinkle formation, while materials like Ag and Al, boasting a Young's modulus of approximately $5 \times 10^{10} \ Pa$, and surface energy around $0.05 J/m^2$ exhibit a critical thickness of around 15 nm. Thus, we can see in SI Fig.11, wrinkles have formed in 15nm Ag and Al films but not on 60nm $As_2Se_3$ film

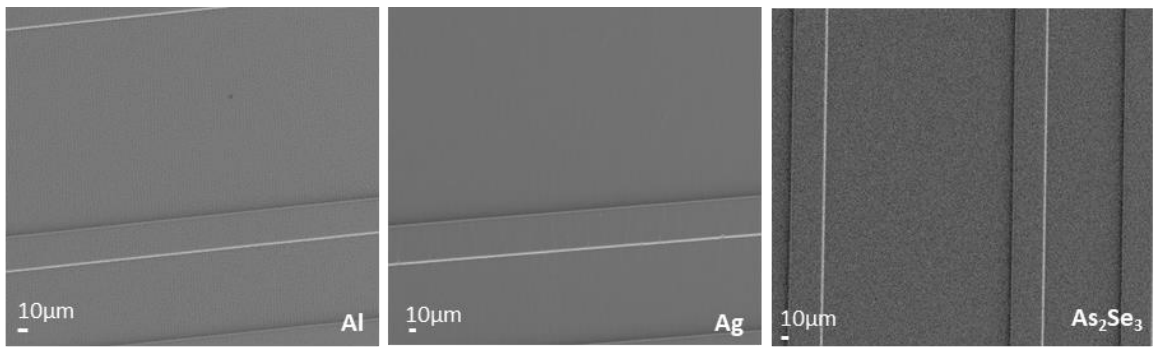

*Figure 11|SEM images showing wrinkle formation in 15nm Al and Ag thin films and the absence of wrinkles in 60nm $As_2Se_3$ thin films. All the thin films were deposited on PDMS5, heated to 80°C for 5 minutes and cooled down to 25°C.*

## 2.4 Prediction from the Wavelength Equation

Our wavelength equation can be efficiently used to find the thermal expansion coefficient of thin films by selecting appropriate values of the different parameters. (Refer Fig. 10 and Fig. 11).

| Material | Film thickness (nm) | Surface energy ($J/m^2$) | Young's Modulus ($Pa$) | $\Delta T(K)$ | Wavelength (µm) | $\alpha_s(K^{-1})$ PDMS 5 | $C_w$ ($\times 10^{-6} m$) | $\alpha_f(K^{-1})$ (predicted from Wavelength Equation) | $\alpha_f(K^{-1})$ (reported in literature) |
|---|---|---|---|---|---|---|---|---|---|
| As$_2$Se$_3$ | 100 | 0.2 | $10^{10}$ | 75 | 3.3 | $216 \times 10^{-6}$ | 3.07 | $22.3 \times 10^{-6}$ | $22 \times 10^{-6}$ |
| Al | 15 | 0.07 | $4.5 \times 10^{10}$ | 55 | 3.7 | $216 \times 10^{-6}$ | 3.5 | $25.51 \times 10^{-6}$ | $18 - 30 \times 10^{-6}$ |
| Ag | 15 | 0.055 | $3.5 \times 10^{10}$ | 55 | 3.2 | $216 \times 10^{-6}$ | 3.15 | $27.57 \times 10^{-6}$ | $19 - 31 \times 10^{-6}$ |

## 2.5 Requirement of a critical length for wrinkle formation and our proposed Critical Length Equation

By combining SI Equation 4 and SI Equation 10, the stress energy can be defined as

$$U = Y_f[(\alpha_s - \alpha_f)\Delta T]^2 whL \quad (13)$$

Thus, in order to overcome the critical stress required for wrinkle formation, the compressive stress should act along a length that is greater than the critical length required for the system. This is why as the distance between the grooves increases, the 1D wrinkles give way to 2D wrinkles because then both the lateral and longitudinal stress are greater than the critical stress required by that system for wrinkle formation (SI Fig. 12).

From SI Equation 7 and SI Equation 11 we can conclude that, if

$$L < C_l \frac{\gamma_f d}{Y_f h[(\alpha_s - \alpha_f)\Delta T]^2}$$

then wrinkle formation does not take place. Here $C_l$ is a unitless proportionality constant and is $= Cn$. Thus, we arrive at the critical length equation which is defined as

$$L_c = C_l \frac{\gamma_f d}{Y_f h[(\alpha_s - \alpha_f)\Delta T]^2} \quad (14)$$

During the formation of 2D wrinkles the lateral and longitudinal stress are not equal. The lateral stress being lesser than the longitudinal stress, the periodicity of the wrinkles due to the lateral stress is lesser.

The proportionality constant $C_l$ depends on the following parameters

i. For a given expanded length $L$, higher the compressive stress higher is the periodicity $d$ and lower is the value of $n$. Since $n$ is directly proportional to $C_l$, a decrease in the value of $n$ also decreases $C_l$ and the critical length $L_c$. Thus, for higher compressive stress the critical length decreases.

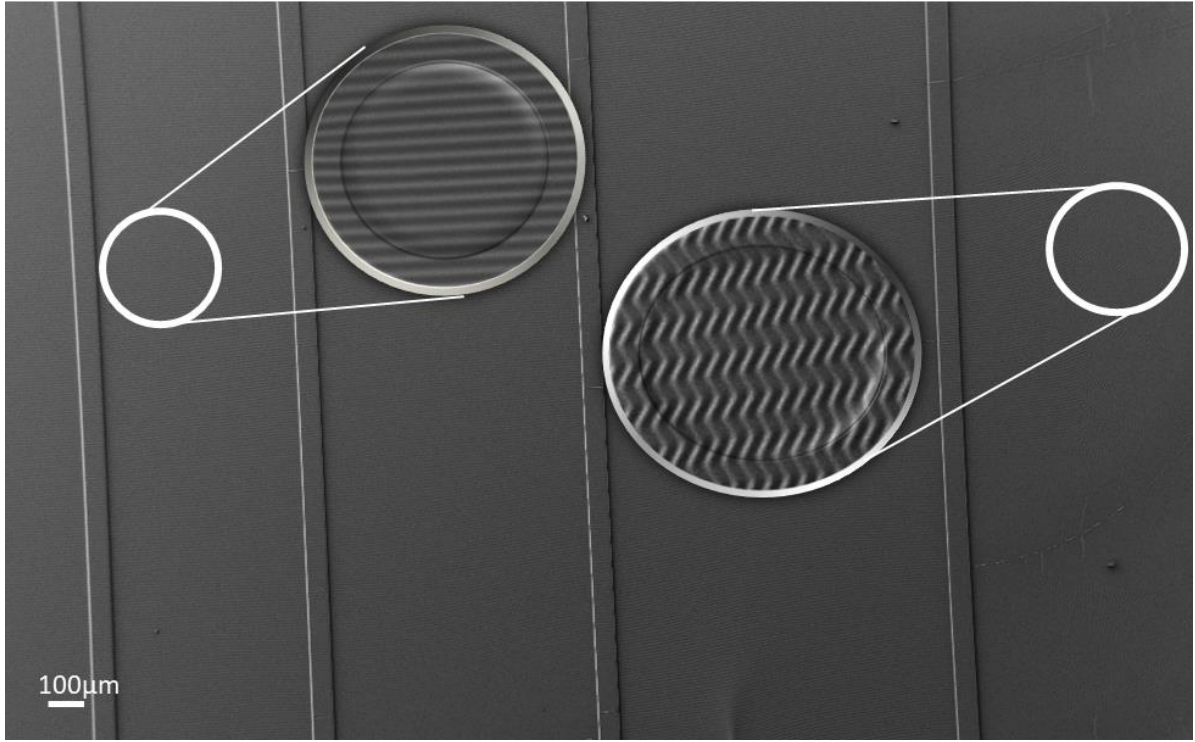

*Figure 12|Large area SEM image of wrinkled As$_2$Se$_3$ thin film on PDMS5 substrate. As the distance between the grooves increases to 1150μm, 2D wrinkles start appearing. Film thickness: 100nm, Heating temperature 120°C.*

## 2.6  Comparison of results from Critical Length Equation with experimental results

If any or all of $\Delta T$, $(\alpha_s - \alpha_f)$ and $h$ increases, then the stress acting on the film also increases, potentially allowing compressive stress along a shorter length to surpass the critical stress necessary for wrinkle formation. This explains why 1D wrinkles form when a 150nm As2Se3 film on PDMS5 is heated to 120°C and then cooled to 25°C, whereas 2D wrinkles emerge when the substrate is changed to PDMS10 while keeping other variables constant (SI Fig. 13a and b). This difference occurs because the thermal expansion coefficient of PDMS10 is higher than that of PDMS5.

Similarly, with increasing $\Delta T$, $L_c$ decreases. For instance, when a 150nm As$_2$Se$_3$ film deposited on PDMS10 is heated to 100°C, it forms 1D wrinkles, while 2D wrinkles appear at 120°C (SI Fig. 13b and c).

A reduction in the surface energy and an increase in the Young's modulus of the thin film material lowers the critical length. Consequently, 1D sinusoidal wrinkles are observed in a 100nm As$_2$Se$_3$ thin film (with $\gamma_f = 0.2 J/m^2$ and $Y_f = 1 \times 10^{10}\ Pa$ ) on a PDMS 5 substrate when heated to 80°C and subsequently cooled, whereas 2D wrinkles are observed in Ag and Al thin films, which have $\gamma_f \approx 0.04 J/m^2$ and $Y_f \approx 4 \times 10^{10}\ Pa$ under similar conditions (SI Fig. 13d, e and f). The critical Length equation has been experimentally found to be more accurate for well-ordered wrinkles, i.e. for wrinkles on As$_2$Se$_3$ thin films.

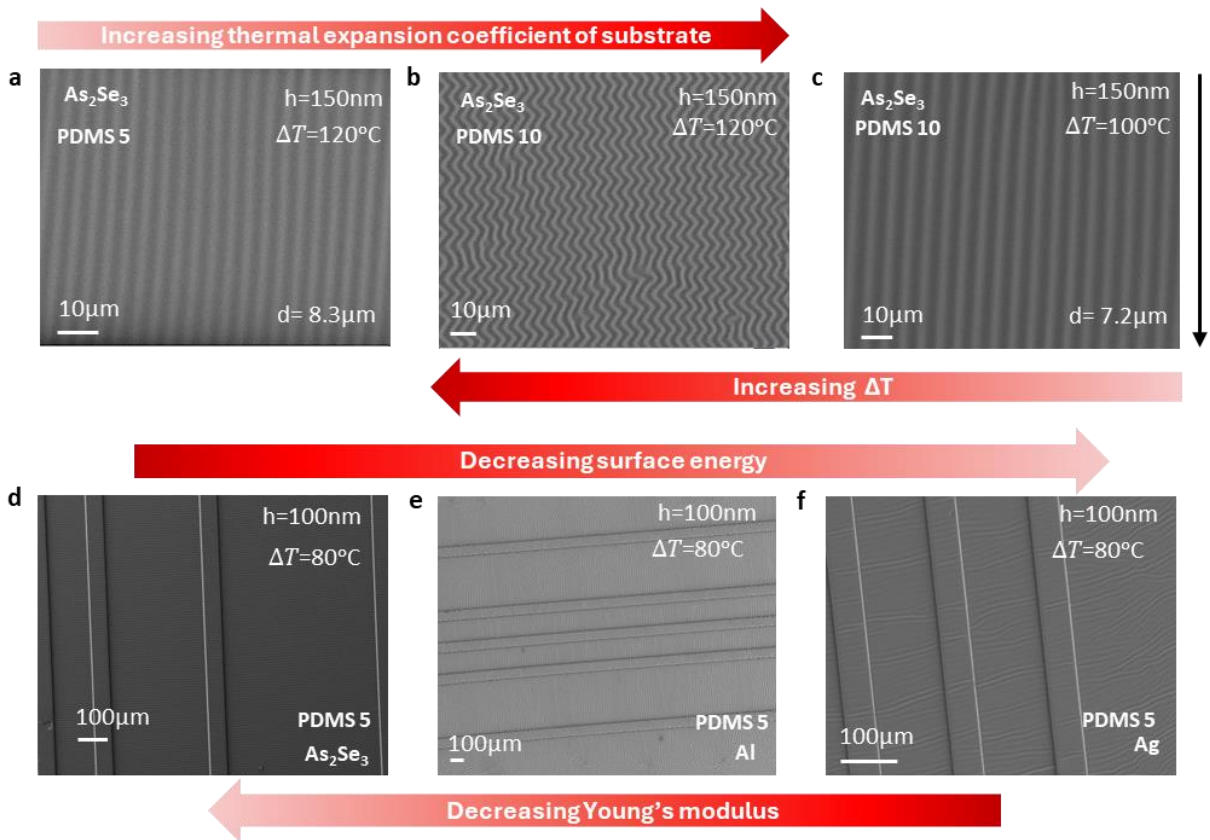

*Figure 13|SEM images a, 150nm As$_2$Se$_3$ deposited on PDMS5, heated to 120°C. b, 150nm As$_2$Se$_3$ deposited on PDMS10, heated to 120°C. c, 150nm As$_2$Se$_3$ deposited on PDMS10, heated to 100°C. a-c, Lateral distance 950nm, black arrow indicates the direction of lateral distance. d, 100nm As$_2$Se$_3$ deposited on PDMS 5, heated to 80°C. e, 100nm Al deposited on PDMS 5, heated to 80°C. f, 100nm Ag deposited on PDMS 5, heated to 80°C. All the samples were cooled down to 25°C for wrinkle formation.*

2D wrinkles or zigzag wrinkles are a superposition of two sets of 1D wrinkles formed along two different directions whereas, 3D wrinkles or labyrinth isotropic wrinkles are a superposition of multiple sets of 1D wrinkles formed along different directions.

### 2.6.1. Effect of heating temperature and substrate softness on wrinkle formation and critical length

As the heating temperature increases, the compressive stress acting on the thin film while it cools, also increases. Wrinkle formation requires the compressive stress acting on the thin film to be greater than its critical stress. The formation of large area, uniform, ordered wrinkles requires a uniform force acting on the film which can be achieved within an optimal temperature range. The lack of wrinkle formation at temperatures much below this range is attributed to the insufficient generation of compressive stress on the film (Figure 14). Conversely, when the system is heated to temperatures exceeding this range by a large extent, disorderedness starts in the thin films (Figure 15).

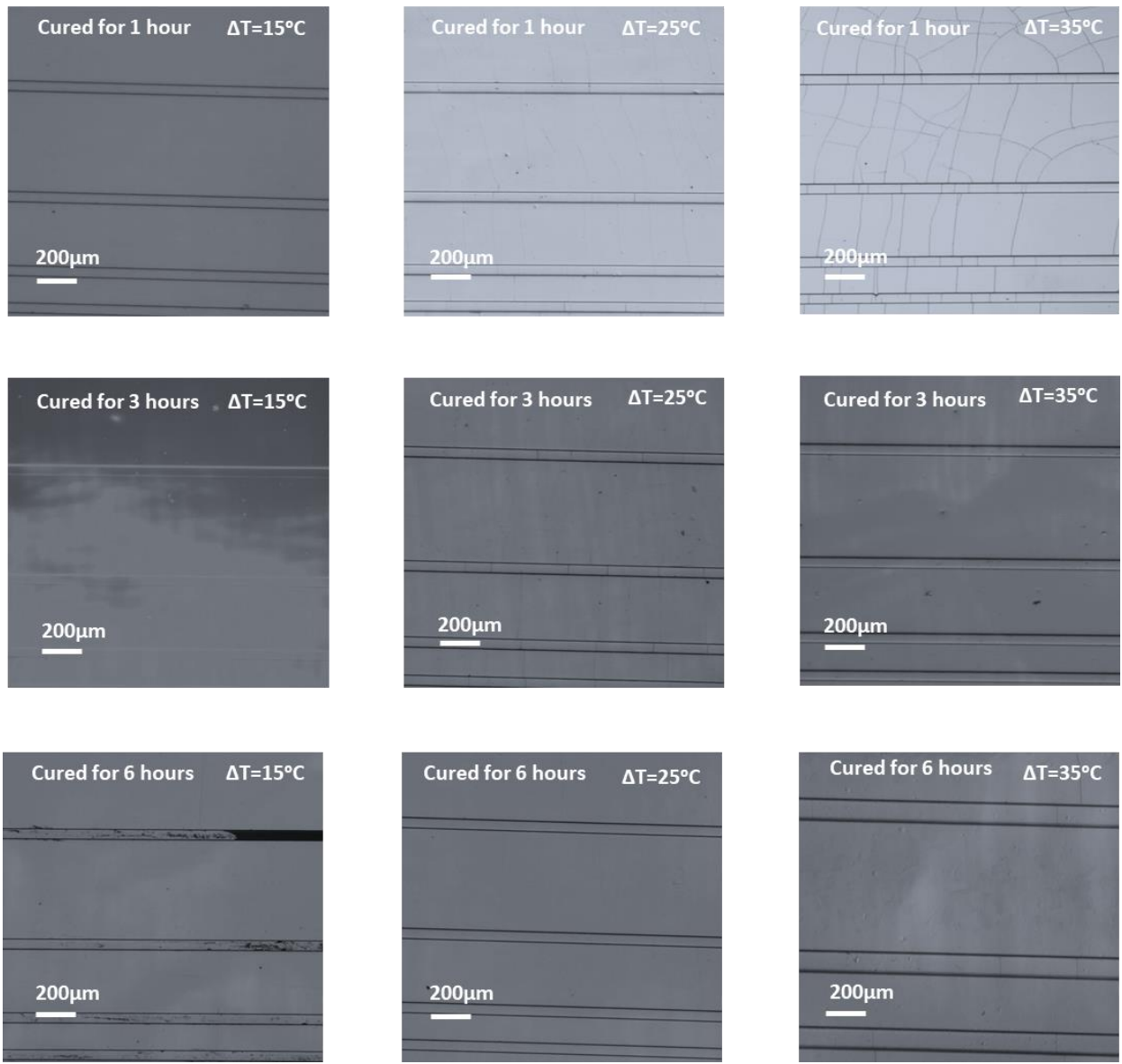

*Figure 14|Optical microscope images illustrating the absence of wrinkles in bilayer systems, consisting of 100nm As$_2$Se$_3$ thin films deposited on PDMS 5 substrates, when heated to temperatures lower than the optimum temperature range for the bilayer system. The PDMS substrates were cured for 1 hour (Top), 3 hours (Middle), and 6 hours (Bottom). The thermal expansion coefficient being higher in PDMS substrates cured for 1 hour, during heating the tensile stress acting over the film deposited atop is also higher leading to the formation of cracks as seen in the Top Right image. For substrates cured for 3 hours and 6 hours, cracks are absent.*

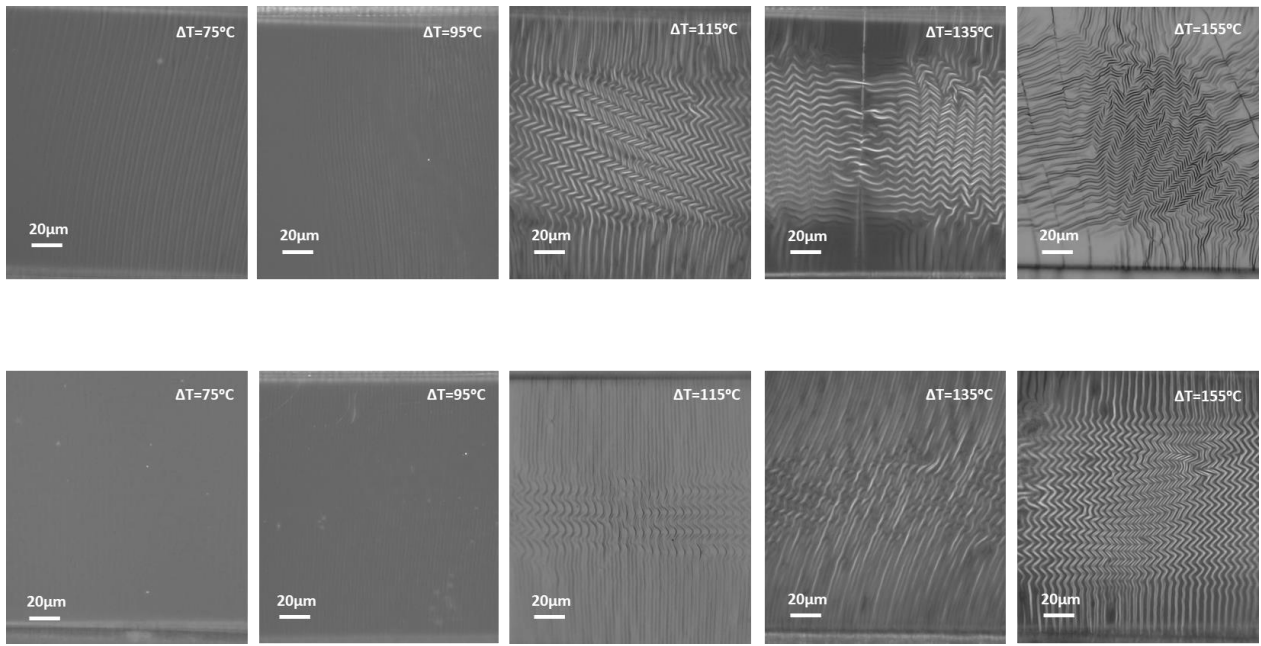

*Figure 15|Optical microscope images illustrating the effect of heating temperature on the orderliness of the wrinkles. Substrate: PDMS 5. Thin film material: 100nm As$_2$Se$_3$. (Top) Substrates cured for 1 hour. (Bottom) Substrates cured for 6 hours. As the heating temperature increases, the 1D ordered wrinkles give way to 2D wrinkles, finally leading to multiple cracks and disordered labyrinthine wrinkles. Again, the wrinkles are more ordered for substrates with higher curing time*

Figure 15 illustrates that as the heating temperature increases, 1D ordered wrinkles give way to 2D wrinkles, finally leading to multiple cracks and disordered labyrinthine wrinkles after cooling down to the room temperature (25°C). As the curing time of the PDMS 5 substrate increases from 1 hour to 6 hours, the wrinkles become more ordered with respect to increase in heating temperature. For $\Delta T = 115°C$, 2D wrinkles have just started forming on PDMS substrates cured for 6 hours, whereas the 2D wrinkles are more prominent on substrates cured for 1 hour. At $\Delta T = 135°C$, cracks have started forming for substrates cured for 1 hour. At $\Delta T = 155°C$, multiple cracks have formed leading to disordered labyrinthine wrinkles on the substrates cured for 1 hour, whereas the substrate cured for 6 hours still has 2D wrinkles, with much lesser number of cracks. Thus, the critical length decreases with increase in heating temperature and duration of curing of PDMS substrate.

## 3. PHYSICS OF WRINKLED SURFACES ACTING AS DIFFRACTION GRATINGS

### 3.1 Origin of diffraction spots from an ordered sinusoidal surface

Diffraction occurs when a wave encounters an obstacle or aperture that causes it to bend around corners or spread out thereby producing periodic alterations in its phase, amplitude or both. The origin of diffraction spots from an ordered sinusoidal surface can be understood through Huygens-Fresnel principle and wave interference. According to this principle, every point on a wavefront can be considered as a source of secondary spherical wavelets. When a wavefront encounters an obstacle, these secondary wavelets interfere with each other constructively and destructively, leading to the formation of a diffraction pattern. An ordered sinusoidal surface acts as a reflection-based diffraction grating because its crests resemble a repetitive array of diffracting elements i.e. obstacles for an incoming wave of light. Each part of the surface acts as a source of secondary wavelets, and the interference between these wavelets results in constructive interference in certain directions and destructive interference in others. The diffraction orders and the spacing between the spots depend on the periodicity of the wrinkles (which is equivalent to the periodicity of the wrinkle crests), the angle of incidence, wavelength of the incident wave and the angle of detection.

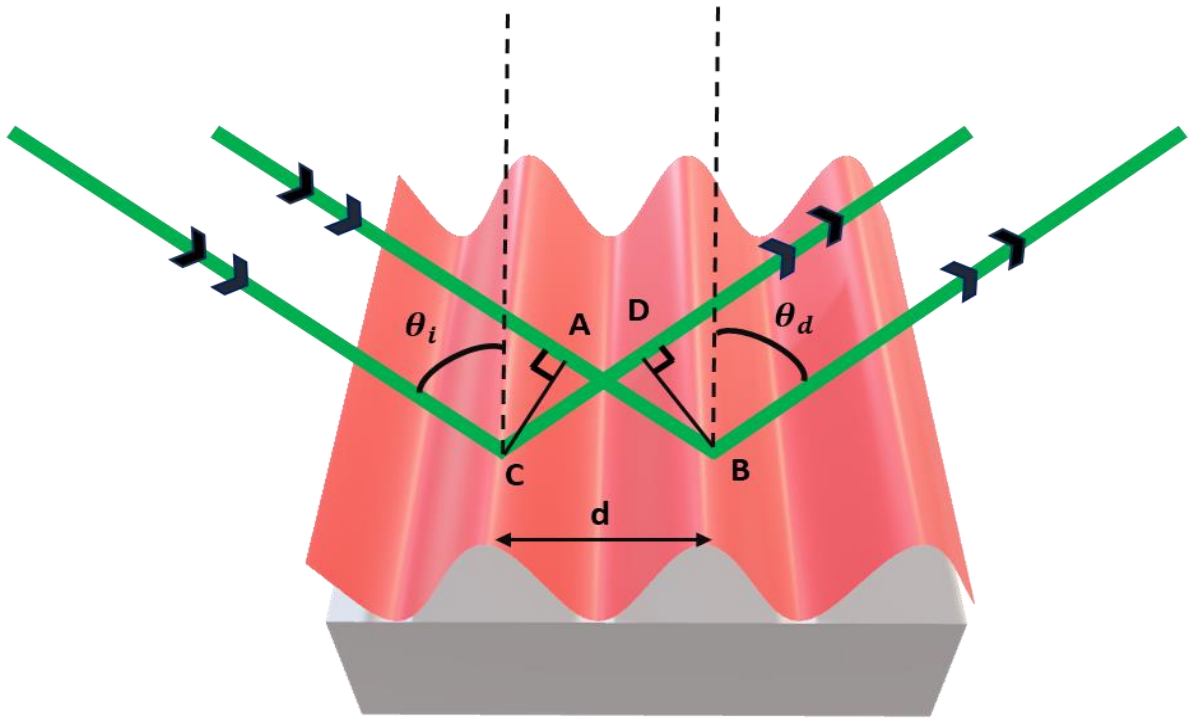

*Figure 16|Schematic of diffraction from a wrinkled surface of periodicity d. Angle of incidence is $\theta_i$, angle of detection of rays is $\theta_d$. If the wavelength of the waves is λ, for constructive interference the path difference between two adjacent waves should be an integral multiple of λ, i.e. $CD - AB = m\lambda$, where m is an integer.*

When plane waves of wavelength λ are incident at an angle $\theta_i$, on a wrinkled surface of periodicity d, in order to get constructive interference at a detection angle $\theta_d$, the following relation must be satisfied

$$CD - AB = m\lambda$$

Where m is an integer and also the diffraction order. Therefore, we get the relation,

$$d \sin \theta_d + d \sin \theta_i = m\lambda$$

$$m\lambda = d(\sin \theta_d + \sin \theta_i) \tag{15}$$

Thus, for a given wavelength (λ) of incident light, incident at an angle $\theta_i$ , we can get different diffraction orders by tuning the periodicity of the wrinkles.

### 3.2   Orientation of diffraction spots

The alignment of the diffraction spots relies on the orientation of the plane of incidence relative to the direction of the wrinkles. When the plane of incidence is oriented parallel to the direction of the wrinkle array, vertical diffraction spots appear. Conversely, if the plane of incidence is perpendicular to the wrinkle array's direction, horizontal diffraction spots are observed.

As we know that the diffraction spots represent the hkl planes in the reciprocal lattice of the Bravais lattice, the ordered wrinkles are analogous to a set of parallel hkl planes. As the orientation of the wrinkles with respect to the incident plane changes, the hkl plane also changes thereby altering the orientation of the diffraction spots. This is why we get circular diffraction patterns for isotropic labyrinth wrinkles. These ring like diffraction patterns find applications in cataract surgeries where circular incisions are needed to be made.

In SI Fig. 17  we can see that as the orientation of the wrinkles with respect to the incident beam plane changes from 0° to 90° in steps of 30°s, the orientation of the diffraction spots also changes.

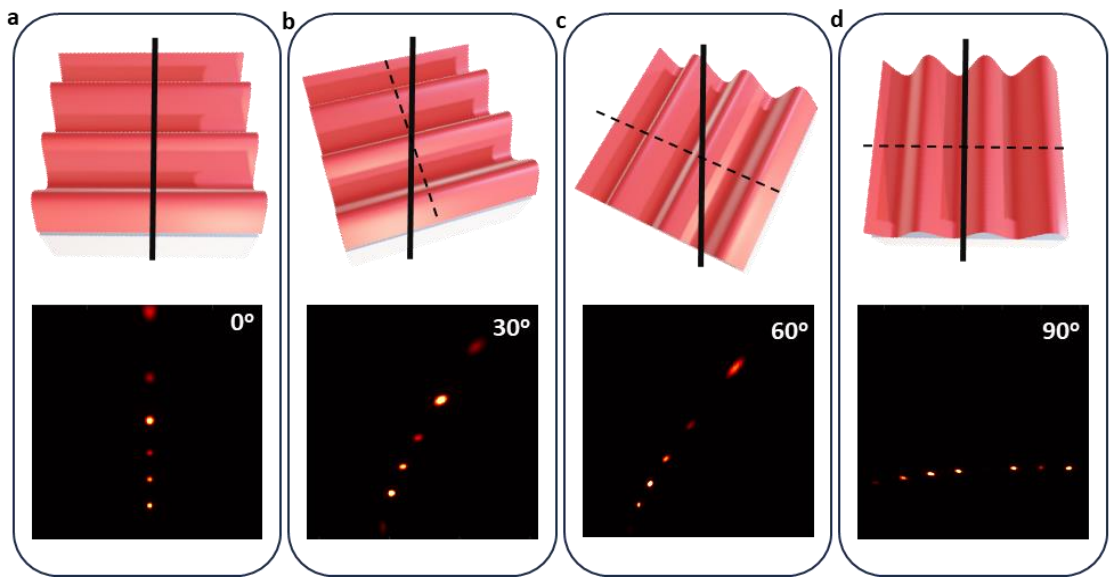

*Figure 17| Schematics depicting the orientation of wrinkles and the corresponding simulation images for the diffraction spots. The simulations were done for a periodicity of 6.2µm, amplitude of 400nm and incident light of wavelength 500nm.*

## 3.3 Wrinkles for optical diffusion

Disordered labyrinthine wrinkles can be considered as superposition of multiple sets of 1D sinusoidal wrinkles at different angles to each other. Thus, when light diffracts over disordered labyrinthine wrinkles, ring like circular diffraction patterns are formed on the detector screen. These wrinkles form when the samples are heated to sufficiently high temperatures followed by subsequent cooling. During cooling as the temperature of the bilayer system decreases, the wavelength and the disorderliness of the wrinkles increases. This subsequent cooling leads to an increase in the number of diffraction rings formed on the screen

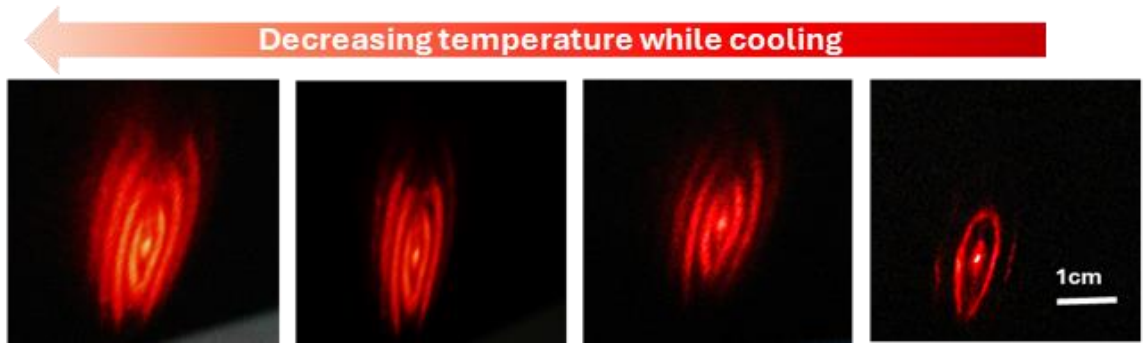

*Figure 18| Experimental images of a tunable optical diffuser, illustrating the dependency of diffraction rings on the cooling temperature*

## 3.4 Disappearance and reappearance of colors during thermal processing

When the wrinkled samples are heated to sufficient temperature the wrinkles disappear due to the thermal expansion in the thin film and also due to the tensile stress acting on it because of the thermal expansion mismatch between the film and the substrate. Diffraction doesn't take place in the absence of wrinkles i.e. on a flat surface, thereby leading to the disappearance of colors on being heated. On subsequent cooling the colors reappear due to the formation of wrinkles which diffract light. The samples are responsive to simple thermal heating as well as joule heating (since both lead to an increase in temperature of the bilayer system), thereby making them excellent candidates for adaptive visible camouflage

## 3.5 Dependence of diffraction orders on temperature

As discussed before, as the heating temperature increases the wavelength(periodicity) of the wrinkles also increases, thereby increasing the number of diffraction orders. For a given wavelength and angle of incidence of incident light, more the periodicity, the more the diffraction orders observed (SI Equation 15).

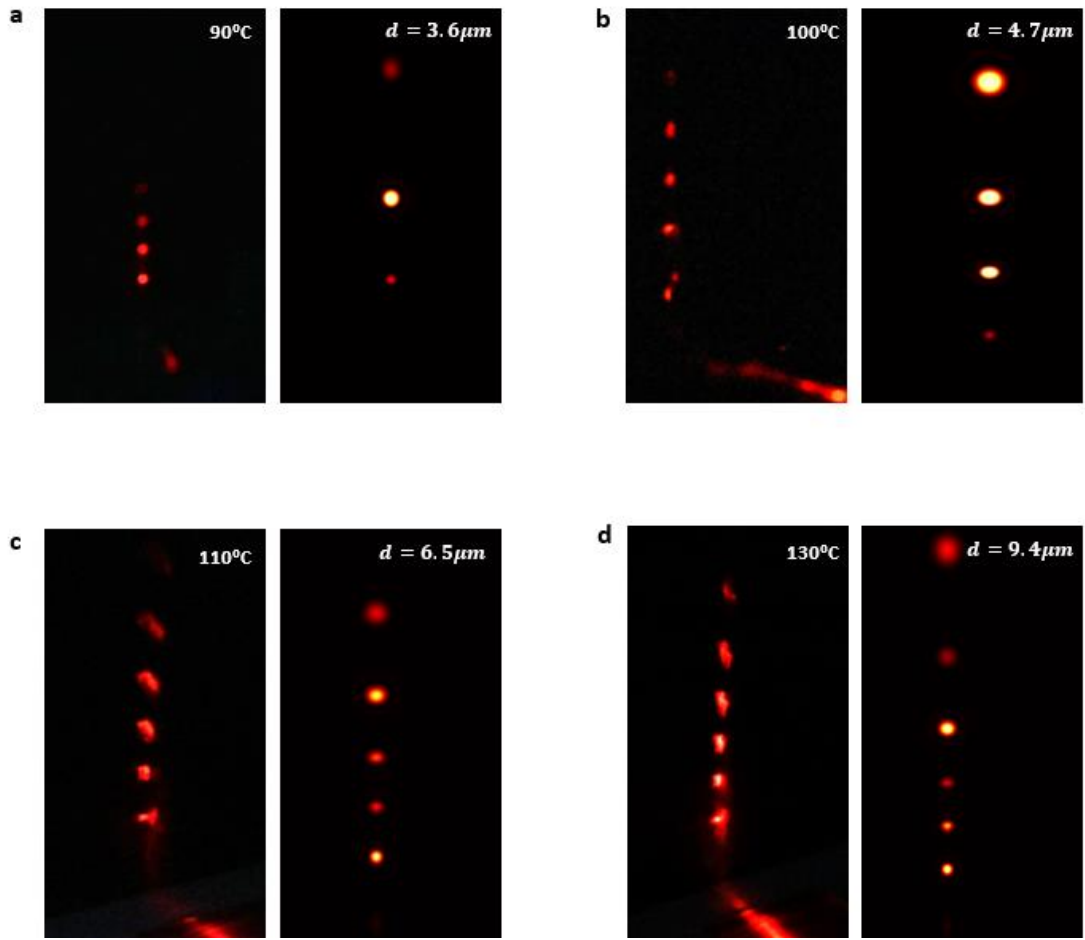

*Figure 19: Experimental images(left) and corresponding simulation images(right). Sample: 150nm $As_2Se_3$ deposited on PDMS 5. a, heated to 90ºC to get wavelength of 3.6µm. b, heated to 100ºC to get wavelength of 4.7µm. c, heated to 110ºC to get wavelength of 6.5µm. d, heated to 130ºC to get wavelength of 9.4µm.*

In SI Fig. 19, the experimental images(left) and the corresponding simulation images (right) illustrate how the number of diffraction orders increases due to the increase in periodicity of the wrinkles when the same sample is heated to different temperatures followed by subsequent cooling.

## 3.6 Simulation of diffraction spots

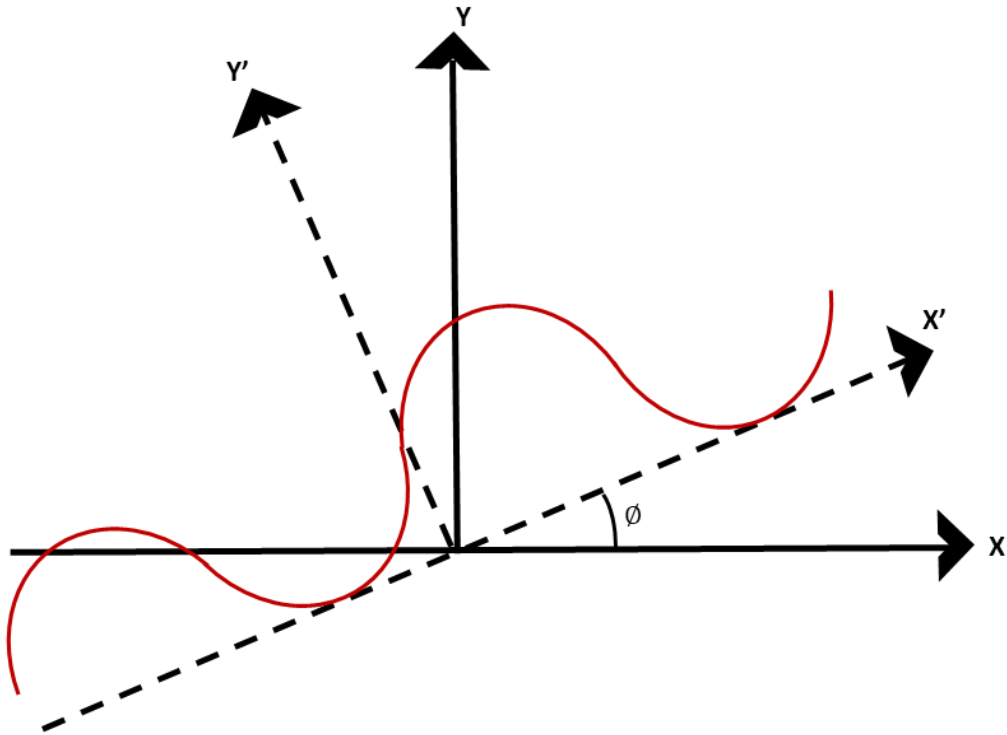

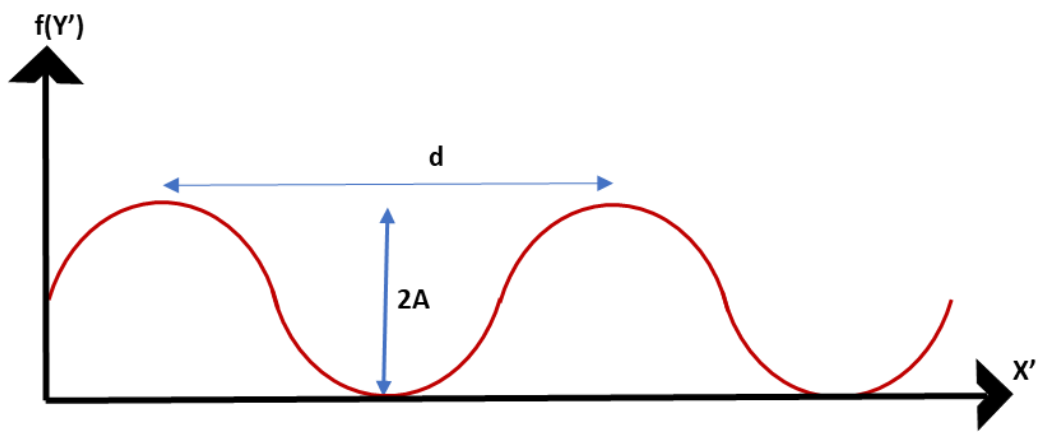

*Figure 20| a, A global coordinate system (X, Y), a local coordinate system (X',Y') rotated by an angle ∅ with respect to global coordinates and a sinusoidal variation along X' axis. b, Cross-sectional image of the sinusoidally varying function along X' axis.*

For the MATLAB simulations we have considered two coordinate systems, a global coordinate system (X, Y) and a local coordinate system (X', Y') rotated by an angle ∅ with respect to the global coordinates and have placed a sinusoidally varying function along the X' axis, as shown in SI Fig.20a. The cross-sectional image shown in SI Fig.20b illustrates that the function of the sinusoidal structure can be defined as,

$$f = A + A\sin(k_d X')$$

Where $k_d = 2\pi/d$ and d is the periodicity of the structure. For normal incidence of light along -Y direction, since the amplitude of the structure is very low as compared to its periodicity, the phase profile on the (X', Y', 2A) plane can be defined as,

$$\emptyset(X',Y') = k_0(2l(X'))  \tag{17}$$

Where $l(X') = 2A - f$

Therefore,
$$l(X') = A - A\sin(k_d X')  \tag{18}$$

The local co-ordinates is related to the global co-ordinates by the rotation matrix and the relation is defined as,

$$\begin{pmatrix} X' \\ Y' \end{pmatrix} = \begin{pmatrix} \cos\emptyset & \sin\emptyset \\ -\sin\emptyset & \cos\emptyset \end{pmatrix} \begin{pmatrix} X \\ Y \end{pmatrix}  \tag{19}$$

From this relation we get,

$$X' = X\cos\emptyset + Y\sin\emptyset  \tag{20}$$

Thus,
$$l(X,Y) = A - A\sin(k_d(X\cos\emptyset + Y\sin\emptyset))  \tag{21}$$

and the phase is defined as

$$\emptyset(X',Y') = k_0 2(A - A\sin(k_d(X\cos\emptyset + Y\sin\emptyset)))  \tag{22}$$

Ignoring the constant phase we get,

$$\emptyset_1(X',Y') = -k_0 2A\sin(k_d(X\cos\emptyset + Y\sin\emptyset))  \tag{23}$$

Considering the light to be incident along X axis, making an angle $\theta$ with respect to Y axis, the additional phase due to this angle of incidence is defined as,

$$\emptyset_2(X,Y) = k_0 X\sin\theta  \tag{24}$$

Therefore, the net phase in the plane (X', Y', 2A) is defined as,

$$\emptyset(X,Y) = \emptyset_1(X,Y) + \emptyset_2(X,Y)  \tag{25}$$

And the corresponding filed is,

$$E(X,Y) = e^{i[k_0 x\sin\theta - 2k_0 A\sin(k_d(X\cos\emptyset + y\sin\emptyset))]}  \tag{26}$$

Considering a finite size of the aperture i.e. limiting the spot size, the field is defined as,

$$E(X,Y) = \begin{cases} e^{i[k_0 x\sin\theta - 2k_0 A\sin(k_d(X\cos\emptyset + y\sin\emptyset))]}, & \sqrt{(X^2 + Y^2)} \leq R \\ 0, & \sqrt{(X^2 + Y^2)} > R \end{cases}  \tag{27}$$

Where R is the radius of the spot.

We have ignored the edge effects since we are in scalar diffraction theory regime as R is very large compared to the periodicity of the sinusoidal structure. Once the field on the Z=2A plane is known, we can calculate the field at any point ($x_0,y_0,z_0$) using the Ray Sommerfeld diffraction formula which is defined as,

$$E(x_0, y_0, z_0) = \iint\limits_{-\infty}^{\infty} \frac{E(X,Y,Z)}{\lambda} \left(\frac{1}{kl_0} - i\right) \frac{(Z - 2A)}{l_0} \frac{e^{ikl_0}}{l_0} dxdy \tag{28}$$

Where $l_0 = \sqrt{(2 - 2A)^2 + (X - x_0)^2 + (Y - y_0)^2}$ and $\lambda$ is the wavelength of the incident monochromatic light. Using SI Equation 28 we calculate the field at a plane Z=1cm (since in our experiments the plane was placed at a distance of 1cm from the sample) to get the diffraction patterns.

## 4. ROLE OF WRINKLES IN INCREASING THE OUT-COUPLING EFFICIENCY

We know that when a ray passes from a medium of higher refractive index to lower refractive index, if the angle of incidence is higher than the critical angle for that pair of media, then total internal reflection (TIR) takes place. TIR reduces the outcoupling efficiency of encapsulated light sources and photovoltaics like solar cells. The presence of wrinkles in the top-most layer increases the critical angle for a pair of media thereby increasing the out-coupling efficiency. This happens because of the change in the position and angle of the normal to the free surface of a film, as illustrated in SI Fig. 18a and b.

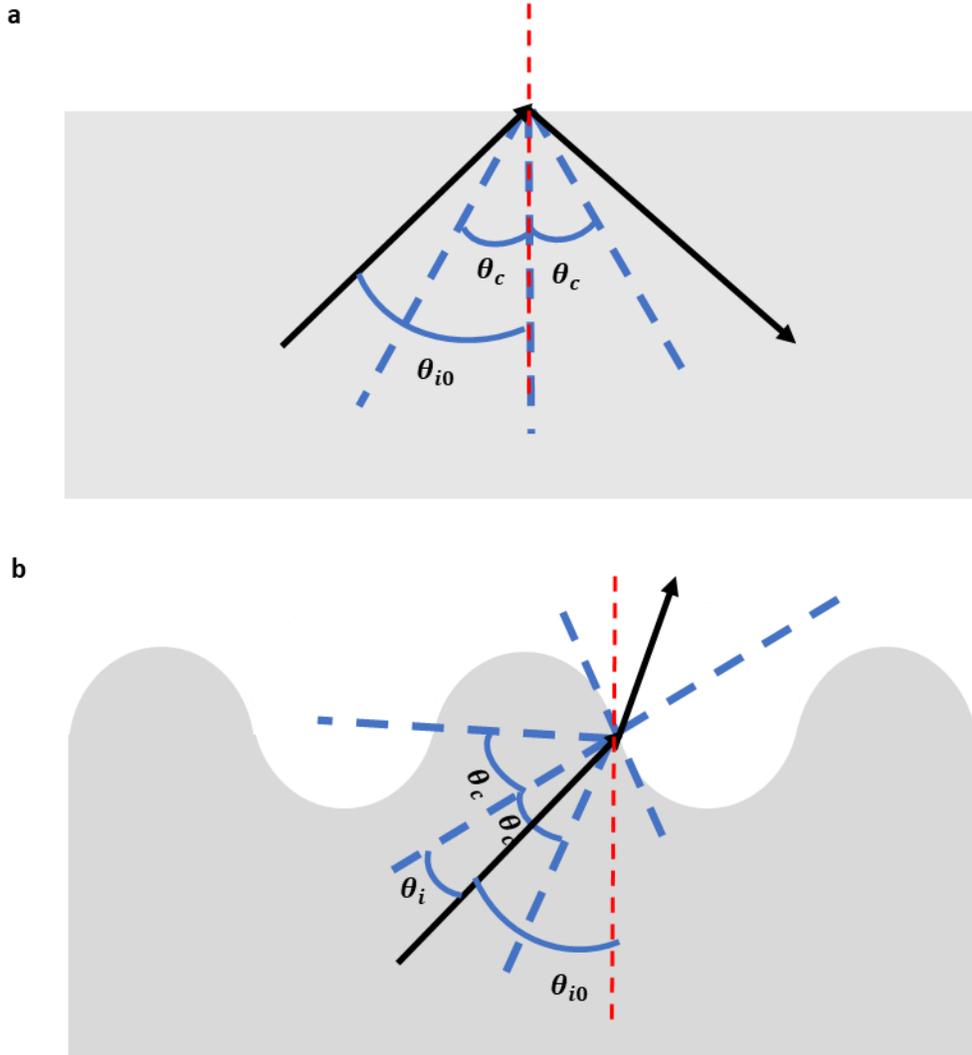

*Figure 21| a, schematic showing TIR when the angle of incidence of a beam going from a higher refractive index to lower refractive index is greater than the critical angle of a given pair of media, i.e. $\theta_{i0} > \theta_c$. b, schematic showing the absence of TIR even when the angle of incidence of a beam going from a higher refractive index to lower refractive index is greater than the critical angle (with respect to the normal to the plane surface) of a given pair of media, i.e. $\theta_i < \theta_c$ even though $\theta_{i0}$ is still the same.*

## 5. NOVELTY OF THE WORK

### 5.1 Novelty in the processing (Ordered, Large area, Defect-free) ,Calibration, Materials and Applications

While wrinkles have been fabricated through various techniques, such as LASER treatment, exposure to UV, visible, or NIR light, plasma treatment, chemical swelling, mechanical strain induction, and thermal annealing, most of them suffer from limitations such as restricted fabrication area, the presence of defects like cracks and delaminations, and disorderliness. Here, we utilize a straightforward thermal processing approach to achieve large area, orderly, and defect-free wrinkles. In this work, we demonstrate, for the first time, a simple calibrated method for the formation of wrinkles of desired periodicity and orientation. Additionally, we have explored various thin film materials like chalcogenides, metals, and polymers, confirming the versatility of our method across these materials.

| Sl No. | Processing for wrinkle formation | Substrate | Thin film material | Defects formed in wrinkles | Calibrated with respect to wrinkle causing stimuli | Type of wrinkles formed | Applications | Reference |
|---|---|---|---|---|---|---|---|---|
| 1 | **Mechanically induced** Formation of thin layer of SiO$_x$ on UVO treated PDMS, Uniaxial stretching of PDMS substrate, Poisson effect causing compressive stress and leading to wrinkle formation in direction perpendicular to the stretching direction | PDMS | - | Cracks | Yes (d as a function of pre-strain) | 1D disordered sinusoidal | Tunable optical transmittance, Smart windows, Switchable elastomeric display | [14] |
| 2 | **Mechanically induced** Uniaxial stretching of PDMS substrate, Sputter-coating with gold, Wrinkle formation upon release of the pre-strain | PDMS | Au | Not reported | Yes (d as a function of pre-strain) | 1D ordered sinusoidal (shown over a 10μm×10μm area) | Micro-strain sensing of rigid substrates | [15] |
| 3 | **Mechanically+ Thermally induced** (i) Sputter-coating of Au film on epoxy at 20°C, Uniaxial stretching of film/substrate system and heating to 65°C, Poisson effect causing compressive stress and leading to wrinkle formation in direction perpendicular to stretching direction (ii) Uniaxial stretching of epoxy substrate at 65°C and subsequent cooling to 20°C, Sputter-coating Au film on pre-stretched substrate at 20°C, Releasing from pre-stretched condition and | epoxy | Au | Cracks | No | 1D disordered | Not shown | [16] |
| | heating to 65°C, leading to a compressive strain in the metal film and subsequent formation of wrinkles | | | | | | | |
| 4 | **Mechanically induced** Uniaxial stretching of PDMS substrate, Ion-coating with gold, Wrinkle formation upon release of the pre-strain | PDMS | Au | Cracks and delaminations | Yes(partially) Wavelength (d) vs. nominal compressive strain plotted | 1D ordered | Not shown | 17 |
| 5 | **Mechanically induced** Spin-coating of polymer films on UV treated Si wafer and placing cured PDMS films on the supported polymer films, Transferring the polymer films onto PDMS substrates by immersing the system in water, Wrinkle formation upon applying mechanical strain on the film/PDMS system | PDMS | PS | Cracks | No | 1D ordered sinusoidal | Measuring elastic moduli of thin polymer films | 18 |
| 6 | **Mechanically induced** Spin-coating of LC polymer films on UV treated Si wafer and placing cured PDMS films on the supported polymer films, Transferring the polymer films onto PDMS substrates by immersing the system in water at room temperature for 10 hours, Achieving wrinkle formation through lateral uniaxial compression on the polymer/PDMS system using a Hoffman type pinchcock | PDMS | Az containing LC polymer | Not reported | No | Partially ordered 1D sinusoidal | Not shown | 19 |
| 7 | **Mechanically+ Thermally induced** Biaxial pre-stretching of air plasma treated and silanized PS substrates, Incubation in APTES solution to form molecular linking layer on the substrate, Pattern-masking and incubating in Au nanoparticle solution to form Au seed layer, Depositing continuous Au film through electroless deposition on seed layer, | PS | Au | Not reported | No | Disordered isotropic folds | SERs based substrates | 20 |
| | Formation of wrinkles on thin film through thermal processing | | | | | | | |
| 8 | **Mechanically induced** Spin-coating PS or PS/Plasticizer films onto polished hydrophilic Si wafers followed by thermal annealing, Transferring the polymer films onto PDMS substrates by immersing the system in water, Achieving wrinkles by applying uniaxial compressive strain using a translation stage | PDMS | PS | Not reported | No | Ordered sinusoidal | Measurement of residual stress in polymer films | 21 |
| 9 | **Mechanically +Chemical Swelling induced** (i) Sequential stretching of oxygen plasma treated PDMS substrate leading to wrinkle formation (ii) DI water swelling of PHEMA films deposited on UV exposed rigid substrates leading to wrinkle formation | PDMS | Composite of PHEMA+ EGDMA | Not reported | Method (i) is calibrated d($\lambda$) as a function of critical strain | 1D and 2D ordered wrinkles, isotropic disordered wrinkles and creases | Tunable wetting and adhesion, tunable open channel micro-fluidics, strain responsive micro-lens arrays | 22 |
| 10 | **Chemical Swelling induced** Controllable swelling of sugar particles infused polymer matrices immersed in water, Deposition of conducting thin films on swollen substrates to obtain wrinkled thin films upon shrinking of the substrates | PDMS, Ecoflex | Au, Graphene | Not reported | No | Isotropic labyrinthine | Soft electronics | 23 |
| 11 | **Chemical Swelling induced** Adsorption of plasma treated PDMS slices in bovice serum albumin followed by adequate swelling in chloroform, Electroless deposition of Ag films on swollen PDMS substrates, Evaporation of chloroform from the PDMS slices leading to wrinkling of Ag films | PDMS | Ag | Not reported | No | Isotropic labyrinthine | Electronic pressure sensor | 24 |
| 12 | **Chemical Swelling induced** Coating and curing liquid PDMS on PVA films deposited on rigid foundations, Peeling off the PVA/PDMS system from the liquid foundation, | PDMS | PVA | Not reported | No | Isotropic labyrinthine | Anti-counterfeit tags, Smart displays, Water indicators, Light diffusers, Antiglare films | 25 |

| # | Method | Substrate | Film | Defects | | Pattern | Application | Ref |
|---|---|---|---|---|---|---|---|---|
| | Preparing 3 types of wrinkles (i) reversible, (ii) one time reversible, (iii) irreversible by proper UV and oxygen exposure, followed by exposure to moisture | | | | | | | |
| 13 | **Chemical Swelling induced** UVO treatment of PS solution spin-coated on Si wafer, Swelling induced surface wrinkling by exposure of PS film to toluene vapours | Si | PS | Not reported | No | Isotropic circular, Spokes, Targets, Labyrinths, Dots | Not shown | 26 |
| 14 | **Chemical Swelling induced** Peeling off cured PDMS stamp from structured Si master, Immersing the PDMS substrates in Aniline solution of required monomer concentration and polymerization at 2-5°C for 20 minutes to form PANI films, Wrinkled and dewrinkled states achieved in PANI film through swelling/deswelling via acid doping and base dedoping | PDMS | PANI | Multiple repetitions of doping/ dedoping causes cracks and delaminations | No | Isotropic labyrinthine, Ordered 1D sinusoidal strips of width 1-10µm | Smart display | 27 |
| 15 | **LASER induced** UV curing of hybrid acryl-amide titania films spin-coated onto rigid substrates, Formation of localized wrinkled strips via localized UV light illumination and successive deposition of thin film material | Si, Silica | Acrylamide-titania | Not reported | No | 1D ordered, 2D array, Disordered isotropic | Gratings | 28 |
| 16 | **LASER induced** Deposition of Chalcogenide thin films on Si substrates through magnetron sputtering, Wrinkle formation through focused pulsed LASER beam irradiation on rotating sample surface | Si | GeTe, $Ge_{30}Te_{70}$, $Ge_2Sb_2Te_5$ | Not reported | No | Disordered isotropic labyrinthine | Anti-counterfeiting tags | 29 |
| 17 | **LASER induced** Wrinkle formation through LASER irradiation | PDMS | - | Cracks | No | Partially ordered 1D sinusoidal | Not shown | 30 |
| 18 | **UV+Thermally induced** Spin coating PAN films onto pigment containing PDMS substrates, | CNT-PDMS, SuO-PDMS | PAN | Not reported | No | Isotropic labyrinthine | Smart displays, Optical diffusers | 31 |
| | Changing thermal expansion coefficient of PAN films through UV exposure, Wrinkle formation through thermal treatment and subsequent cooling | | | | | | | |
| 19 | **UV induced** Spin coating PAN-BA films onto PDMS substrates, Generation of wrinkles by controlling the modulus of photosensitive PAN-BA layers spatially and temporarily via sequential exposure of UV/NIR light | CNT-PDMS | PAN-BA | Not reported | Yes | Ordered 1D and 2D sinusoidal strips | Generation of dynamic diffraction patterns | 32 |
| 20 | **UV+Thermally induced** Spin-coating PMMA on PDMS substrates followed by heating and cooling, Formation of wrinkle patterns under UV radiation or thermal stimulation of PMMA films | PDMS | PMMA | No | No | Disordered labyrinthine | Tunable UV/thermal sensitive optical diffusers | 33 |
| 21 | **Thermally induced** Heating of multilayer system to a temperature higher than the $T_g$ of second layer and lower than that of the uppermost layer, Formation of wrinkles through compressive stress generation | Si | PS/Al/P4VP | No | No | Isotropic disordered labyrinthine | Estimation of $T_g$ of polymeric thin films | 34 |
| 22 | **Thermally induced** Thermal evaporation of Al thin films on PS films spin-coated on Si wafers, Wrinkle formation on Al films through thermal annealing for atleast 3600s at temperatures above the glass transition temperature of PS | Si | PS/Al | Not reported | No | Isotropic disordered | Estimation of stress-relaxation modulus of polymer thin films confined by both a substrate and a superstrate | 35 |
| 23 | **Thermally induced** Supersonic cluster beam implantation of neutral Au nanoparticles into thermally retractable PS sheets, Wrinkle formation through thermal shrinking of the material at 160°C for 6 minutes | PS sheet | Au nanoparticles | Not reported | No | Isotropic, Disordered 3D,2D and 1D. | Not shown | 36 |
| 24 | **Thermally induced** Spin-coating PAN solution on CNT-PDMS | CNT-PDMS | PAN interlayer, | Not reported | No | 1D,2D ordered | Smart displays, devices for | 37 |
| | substrate followed by sequential exposure to UV light through photomask, Deposition of metal films on PAN layer through e-beam evaporation, Wrinkle formation in metal films through heat treatment | | Functional metallic layer of Al or Ge or Au | | | sinusoidal strips | resolving light pollution | |
| 25 | **Thermally induced** Dip-coating borosilicate glass substrate in ZnO sol gel, Heat treatment at 150°C in IR chamber of dip coater, Post annealing in air at 450°C for 1 hour, Multi layers prepared through the same procedure leading to wrinkle formation | Boro-silicate glass | ZnO | No | No | Isotropic labyrinthine | Enhancing photocatalytic degradation of organic pollutants in water | 38 |
| 26 | **Thermally + Mechanically induced** Programming Shape Memory Polymer (SMP) substrates by heating above $T_g$ and stretching along one direction, Consequent shrinking along other two directions due to Poisson effect, Cooling down the SMP to room temperature and coating the surface with thin metallic film, Reheating the bilayer to temperatures above $T_g$ and consequent wrinkling of the thin film | Epoxy resin (EPON826 + D230) | Al | Cracks | No | Partially ordered 1D sinusoidal | Reflectors in organic photovoltaics | 39 |
| 27 | **Thermally induced** Spin-coating PDMS on glass followed by partial pre-curing, Thermal evaporation of Al thin films on PDMS, Enhancement of cross-linking in the PDMS layer due to the temperature of the film leading to freezing of patterns on the PDMS substrate and consequent wrinkling in the thin film | PDMS | Al | Not reported | No | Disordered labyrinthine | Fabrication of surfaces with controlled wettability and reflectance | 40 |
| 28 | **Thermally induced** Spin-coating PS solution on Si substrate, Deposition of thin Al layer through thermal evaporation, Wrinkle formation by heating samples to | PS | Al | Not reported | No | Disordered labyrinthine | Not shown | 41 |
| | 140°C or 170°C for different durations of time | | | | | | | |
| 29 | This paper: **Thermally induced** Spin-coating PDMS solution on patterned Si mold followed by curing at 80°C for 1 hour, Peeling off the PDMS substrate and thin film deposition on its surface, Wrinkle formation in thin films through thermal heating for 3 minutes and subsequent cooling | PDMS 5, PDMS 10 | $As_2Se_3$, Al, Ag, PMMA | No | Yes (Periodicity d as a function of Change in temperature ΔT, also an equation illustrating the critical length required for ordered wrinkle formation in case of thermally induced wrinkles) | Ordered 1D sinusoidal, Ordered 2D sinusoidal, Disordered labyrinthine | Tunable diffraction grating, tunable optical diffusers, devices for increasing the out-coupling efficiency of encapsulated light sources and photovoltaics, Reflective displays and Adaptive visible camouflage, Mechanochromic sensors | Novelty: (i) Calibration of the process in terms of the wrinkle causing stimuli i.e. ΔT (ii)Obtained large area, defect free, ordered and reversible wrinkles through a simple thermal processing technique |

*Table 2|Comparison of our work with existing works on wrinkles*

**The salient novelty accompanying the fabrication process, materials used, calibration of the process and photonic applications, summarized from above discussion are as follows:**

1)**Simple fabrication process:** The process involves deposition of thin films on PDMS substrates followed by simple thermal heating and cooling, thus marking a departure from the previous reliance on intricate chemical, mechanical and lithographic methods and making it highly scalable and cost effective.

2)**Defect-free structures:** The process has been thoroughly studied and engineered to get large area samples that are almost free of cracks.

3)**Calibrated process:** For the first time, this thermal process has been calibrated based on the wrinkle-inducing factors such as temperature change (ΔT), material properties like thermal expansion coefficients, and surface energy. This has led to the development of an equation that aids in achieving customized wrinkle patterns using different materials like Chalcogenides, metals and polymers.

4)**Applications in soft photonics:** The samples developed through this process can find applications in large area$(\sim 100 cm^2)$ tunable reflective displays, smart displays, mechanochromic sensors, temperature controlled and dynamically tunable diffraction gratings and optical diffusers.

# 6. References

1. Müller, A., Wapler, M. C. & Wallrabe, U. A quick and accurate method to determine the Poisson's ratio and the coefficient of thermal expansion of PDMS. *Soft Matter* **15**, 779–784 (2019).

2. Zmrhalová, Z., Pilný, P., Svoboda, R., Shánělová, J. & Málek, J. Thermal properties and viscous flow behavior of As2Se3 glass. *J Alloys Compd* **655**, 220–228 (2016).

3. Málek, J. & Shánělová, J. Structural relaxation of As2Se3 glass and viscosity of supercooled liquid. *J Non Cryst Solids* **351**, 3458–3467 (2005).

4. Shchurova, T. N. & Savchenko, N. D. *Defense Technical Information Center Compilation Part Notice CORRELATION BETWEEN MECHANICAL PARAMETERS FOR AMORPHOUS CHALCOGENIDE FILMS*. *Journal of Optoelectronics and Advanced Materials* vol. 3 (2001).

5. Mel'Nichenko, T. D. *et al.* On the approximate estimation of the surface tension of chalcogenide glass melts. *Glass Physics and Chemistry* **35**, 32–42 (2009).

6. Fang, W. & Lo, C.-Y. *On the Thermal Expansion Coefficients of Thin Films*. *Sensors and Actuators* vol. 84 www.elsevier.nlrlocatersna (2000).

7. Samad, M. I. A. *et al.* Aluminium Thin Film Surface Modification via Low-Pressure and Atmospheric-Pressure Argon Plasma Exposure. *Journal of Surface Investigation* **16**, 421–426 (2022).

8. Lim, Y. Y., Chaudhri, M. M. & Enomoto, Y. Accurate determination of the mechanical properties of thin aluminum films deposited on sapphire flats using nanoindentations. *J Mater Res* **14**, 2314–2327 (1999).

9. Zoo, Y., Adams, D., Mayer, J. W. & Alford, T. L. Investigation of coefficient of thermal expansion of silver thin film on different substrates using X-ray diffraction. *Thin Solid Films* **513**, 170–174 (2006).

10. Oje, A. I., Ogwu, A. A., Mirzaeian, M., Oje, A. M. & Tsendzughul, N. Silver thin film electrodes for supercapacitor application. *Appl Surf Sci* **488**, 142–150 (2019).

11. Laugier, M. *Letter Determination of Young's Modulus in Vacuum-Evaporated Thin Films of Aluminium and Silver*. *Thin Solid Films* vol. 75 (1981).

12. Han, G., Huan, S., Han, J., Zhang, Z. & Wu, Q. Effect of acid hydrolysis conditions on the properties of cellulose nanoparticle-reinforced polymethylmethacrylate composites. *Materials* **7**, 16–29 (2014).

13. Mounder Kouicem, M. *et al.* An investigation of adhesion mechanisms between plasma-treated PMMA support and aluminum thin films deposited by PVD. (2021) doi:10.1016/j.apsusc.2021.150322ï.

14. Li, Z. *et al.* Harnessing Surface Wrinkling–Cracking Patterns for Tunable Optical Transmittance. *Adv Opt Mater* **5**, (2017).

15. Ma, T. *et al*. Micro-strain sensing using wrinkled stiff thin films on soft substrates as tunable optical grating. *Opt Express* **21**, 11994 (2013).

16. Li, J., An, Y., Huang, R., Jiang, H. & Xie, T. Unique aspects of a shape memory polymer as the substrate for surface wrinkling. *ACS Appl Mater Interfaces* **4**, 598–603 (2012).

17. Sun, J. Y., Xia, S., Moon, M. W., Oh, K. H. & Kim, K. S. Folding wrinkles of a thin stiff layer on a soft substrate. in *Proceedings of the Royal Society A: Mathematical, Physical and Engineering Sciences* vol. 468 932–953 (Royal Society, 2012).

18. Stafford, C. M. *et al*. A buckling-based metrology for measuring the elastic moduli of polymeric thin films. *Nat Mater* **3**, 545–550 (2004).

19. Takeshima, T. *et al*. Photoresponsive Surface Wrinkle Morphologies in Liquid Crystalline Polymer Films. *Macromolecules* **48**, 6378–6384 (2015).

20. Gabardo, C. M. *et al*. Rapid prototyping of all-solution-processed multi-lengthscale electrodes using polymer-induced thin film wrinkling. *Sci Rep* **7**, (2017).

21. Chung, J. Y., Chastek, T. Q., Fasolka, M. J., Ro, H. W. & Stafford, C. M. Quantifying residual stress in nanoscale thin polymer films via surface wrinkling. *ACS Nano* **3**, 844–852 (2009).

22. Yang, S., Khare, K. & Lin, P. C. Harnessing surface wrinkle patterns in soft matter. *Adv Funct Mater* **20**, 2550–2564 (2010).

23. Yang, Y. & Zhao, H. Water-induced polymer swelling and its application in soft electronics. *Appl Surf Sci* **577**, (2022).

24. Gao, N., Zhang, X., Liao, S., Jia, H. & Wang, Y. Polymer Swelling Induced Conductive Wrinkles for an Ultrasensitive Pressure Sensor. *ACS Macro Lett* **5**, 823–827 (2016).

25. Zeng, S. *et al*. Moisture-Responsive Wrinkling Surfaces with Tunable Dynamics. *Advanced Materials* **29**, (2017).

26. Chung, J. Y., Nolte, A. J. & Stafford, C. M. Diffusion-controlled, self-organized growth of symmetric wrinkling patterns. *Advanced Materials* **21**, 1358–1362 (2009).

27. Xie, J., Han, X., Zong, C., Ji, H. & Lu, C. Large-area patterning of polyaniline film based on in situ self-wrinkling and its reversible doping/dedoping tunability. *Macromolecules* **48**, 663–671 (2015).

28. Takahashi, M. *et al*. Photoinduced formation of wrinkled microstructures with long-range order in thin oxide films. *Advanced Materials* **19**, 4343–4346 (2007).

29. Martinez, P. *et al*. Laser Generation of Sub-Micrometer Wrinkles in a Chalcogenide Glass Film as Physical Unclonable Functions. *Advanced Materials* **32**, (2020).

30. Qi, L. *et al*. Writing Wrinkles on Poly(dimethylsiloxane) (PDMS) by Surface Oxidation with a $CO_2$ Laser Engraver. *ACS Appl Mater Interfaces* **10**, 4295–4304 (2018).

31. Ma, T. *et al.* Light-driven dynamic surface wrinkles for adaptive visible camouflage. doi:10.1073/pnas.2114345118/-/DCSupplemental.

32. Zhou, L. *et al.* Regulating surface wrinkles using light. *Natl Sci Rev* **7**, 1247–1257 (2020).

33. Jiang, S., Yin, X., Bai, J., Yu, B. & Qian, L. Fabrication of ultraviolet/thermal-sensitive PMMA/PDMS wrinkle structures and the demonstration as smart optical diffusers. *Ceram Int* **49**, 10787–10794 (2023).

34. Yoo, S. S. *et al.* Cumulative energy analysis of thermally-induced surface wrinkling of heterogeneously multilayered thin films. *Soft Matter* **14**, 704–710 (2018).

35. Chan, E. P., Kundu, S., Lin, Q. & Stafford, C. M. Quantifying the stress relaxation modulus of polymer thin films via thermal wrinkling. *ACS Appl Mater Interfaces* **3**, 331–338 (2011).

36. Greco, F. *et al.* Conducting shrinkable nanocomposite based on au-nanoparticle implanted plastic sheet: Tunable thermally induced surface wrinkling. *ACS Appl Mater Interfaces* **7**, 7060–7065 (2015).

37. Chen, S. *et al.* Dynamic metal patterns of wrinkles based on photosensitive layers. *Sci Bull (Beijing)* **67**, 2186–2195 (2022).

38. Daher, E. A., Riachi, B., Chamoun, J., Laberty-Robert, C. & Hamd, W. New approach for designing wrinkled and porous ZnO thin films for photocatalytic applications. *Colloids Surf A Physicochem Eng Asp* **658**, (2023).

39. Chen, Z., Young Kim, Y. & Krishnaswamy, S. Anisotropic wrinkle formation on shape memory polymer substrates. *J Appl Phys* **112**, (2012).

40. Das, A., Banerji, A. & Mukherjee, R. Programming Feature Size in the Thermal Wrinkling of Metal Polymer Bilayer by Modulating Substrate Viscoelasticity. *ACS Appl Mater Interfaces* **9**, 23255–23262 (2017).

41. Yoo, P. J. & Lee, H. H. Evolution of a stress-driven pattern in thin bilayer films: Spinodal wrinkling. *Phys Rev Lett* **91**, (2003).